\documentclass[preprint,showpacs,preprintnumbers,amsmath,amssymb,superscriptaddress]{revtex4-2}

\usepackage{graphicx,graphics}   
\usepackage{graphicx}
\usepackage[utf8x]{inputenc}
\usepackage{color}
\usepackage{subfig}
\usepackage{multirow}
\usepackage{slashed}
\usepackage{ulem}
\usepackage{slashed}
\usepackage{graphicx,amsfonts}
\usepackage{epsfig}

\usepackage{dcolumn}
\usepackage{bm}
\begin{document}
	
	
\title{Challenges for the 3-3-1 models
}

\author{V. Pleitez }%
\email{v.pleitez@unesp.br; \textrm{https://orcid.org/0000-0001-5279-8438}.}
\affiliation{
	Instituto  de F\'\i sica Te\'orica--Universidade Estadual Paulista \\
	R. Dr. Bento Teobaldo Ferraz 271, Barra Funda\\ S\~ao Paulo - SP, 01140-070,
	Brazil
}

	
\date{06/09/2022
}
	
\begin{abstract}
The so-called 3-3-1 models have been widely studied in the literature. However, in almost all cases, as with other models, just simplified versions have been considered. 
On the other hand, the LHC has not neither confirmed nor rejected any model. 
In this paper we discuss some aspects of the 3-3-1 models that must be investigated, or be re-analyzed, firstly theoretically and, eventually, experimentally. 
Some of these issues are independent of the representation content of the particular model, that is, they are independent of the value of the $\beta$ parameter. However, others strongly depend on the particular model considered. 
\end{abstract}

	
	
\maketitle
	
\section{Introduction}
\label{sec:intro}

It is very well known that there are questions which are still open in the context of the standard model (SM), and they are ranging from neutrinos physics,  astrophysics, and cosmology. Just to mention the oldest ones: dark matter~\cite{dm} (however see \cite{Skordis:2020eui}) and dark energy~\cite{DeRujula:2021vvx}, neutrino masses~\cite{ParticleDataGroup:2020ssz} and matter-antimatter asymmetry in the universe~\cite{Planck:2018vyg}. All those issues have been already observed but not explained. There are some possible anomalies i.e., observables that are not in total agreement with the SM predictions, for instance, muon $(g-2)_\mu$~\cite{Abi:2021gix}, $B$-decays~\cite{Aaij:2019wad}, the Cabibbo angle anomaly~\cite{Belfatto:2019swo}, just to mention some of them. Although it is not clear that these anomalies will be confirmed in the near future, we have to pay attention to them~\cite{London:2021lfn}.

Most proposed solutions to explain the possible deviations from the predictions of the SM have been made independently of the model, or using effective operators and, usually each one targeting a specific problem. Moreover, if a new sort of particle is found the respective experiment will not be able, in a first moment,  discern between different models that have this sort of particle. For instance, in searching an extra singly charged vector boson $W^\prime$, it is assumed that the main decay is $W^\prime\to t\bar{b}$~\cite{CMS:2021mux}. However, in some models the decay is $W^\prime\to \bar{t}X$ or $W^\prime\to bY$, where $X$ and $Y$ may be quarks with exotic electric charge, say $5/3$ or $-4/3$. 

It is also important to emphasize that the phenomenological analyses which are inspired in a particular model, usually are carried out without taking into account the matrices (unitary or not) present in all vertices of the model. Although this is easy to understand because the calculations are already complicated, we cannot forget that the SM would be ruled out if it did not have the possibility of accomodate the CKM matrix. 
In the same way, no model should be discarded just by placing constraints in the mass of only one of the particles present in the model and a few more parameters. The usual approaches are also justified since no new particle has been discovered until now. However, this scheme produces a patchwork of possible solutions over distinct issues lacking unity and consistency themselves.
Nevertheless, sooner or later, the experimental data will have to be confronted with at least one specific model, which will be renormalizable and free of anomalies. Or, maybe a new paradigm in elementary particle physics would arise, probably one which will not based in gauge symmetries.

Among all the extensions of the SM proposed in the literature,  we do not know yet if one of them (if any) is realized in nature. However, it is still possible that new physics be confirmed at energies of some TeVs. Some possibilities are the following:
\begin{itemize}
\item Multi-Higgs extension of the SM~\cite{Maniatis:2015gma,Gomez-Bock:2021uyu},
\item Left-right (symmetric or not) models~\cite{Senjanovic:2017ldw,Senjanovic:2019moe,Chavez:2019yal,Diaz:2020pwh},
\item 3-3-1 models~\cite{Singer:1980sw,Pisano:1991ee,Foot:1992rh,Frampton:1992wt,Pleitez:1992xh}
\item A combination of the previous ones~\cite{Reig:2016tuk,Franco:2016hip},
\item Grand unification~\cite{Croon:2019kpe},
\item Supersymmetric version of one of the previous ones~\cite{Montero:2000ng,Rodriguez:2009cd,Babu:2020ncc},	
\item \textbf{None of the above}.
\end{itemize}
In fact, it is still possible, as pointed out above, that the SM is the last model based on gauge principle, and a new paradigm would arise in the physics of the elementary particles and their interactions. Here, of course, we will not consider this possibility. 
In fact, we will focus only on  models in which the gauge symmetries are $SU(3)_C\otimes SU(3)_L\otimes U(1)_X$, called 3-3-1 models, with different representation contents. All of them are free of anomalies only when the three generations are considered at once, and not generation per generation as is the case in the SM. Due to the representation content, all these models have a lot of parameters and we will consider here their origin and why they, sooner or later, they will have to be consider when studying their phenomenology. 

The text is organized as follows. In Sec.~\ref{sec:themodels} we consider general features of these models, for example the motivation to extend the SM symmetry to 3-3-1 is considered in~Sec.~\ref{subsec:321331}; in Sec.~\ref{subsec:qo} we define the electric charge operator and the parameter $\beta$ (which allow us to distinguish different models); in Sec.~\ref{subsec:2331models} we consider two of the most studied 3-3-1 models, one with $\beta=-\sqrt{3}$ in Sec.~\ref{subsubsec:m331} and the other with $\beta=-1/\sqrt{3}$ in Sec.~\ref{subsubsec:nur331}; in Sec.~\ref{subsec:dbeta} we show that models with the same $\beta$ may have different phenomenology, i.e., they are different models because they have different representation contents. The main predictions of the models are considered in Sec~\ref{subsec:predictions}. 
In Sec.~\ref{sec:smp} we show the representation content of the model with $\beta=-\sqrt{3}$ projected onto the symmetry of the SM, includes many extensions of the SM with several scalar multiplets, and also new quarks and leptons which are singlets under the SM symmetries. In Sec.~\ref{sec:interlude} we discuss some of the old models with the same gauge symmetries arguing why they cannot be consider of the 3-3-1 type. 
Sec.~\ref{sec:challenges} is devoted to different challenges that, in our opinion, will have to be confronted with this type of models, but of course, not only with them. The origin of the parameters of the m331 model is considered in Sec.~\ref{sec:parameters}, for quarks and leptons in Secs.~\ref{subsec:quarks} and \ref{subsec:leptons}, respectively.
The challenges in the quark sector are discussed with more detail in Sec.~\ref{sec:chaquarks}; the direct search for exotics quarks in Subec.~\ref{subsec:exotics} and their indirect effect in the possible existence of exotic hadrons in Subsec.~\ref{subsec:ebaryons}.

Sec.~\ref{sec:chaleptons} is devoted to the challenges in the lepton sector,  neutrino masses in Sec.~\ref{subsec:nusmasses}; in 
Sec.~\ref{subsec:rare} we consider rare processes (not confirmed) that motivated the old models with the same gauge symmetry but with different representation content; in Sec.~\ref{subsec:nusexp} we consider some, at present, academic issues in neutrino experiments. In Sec.~\ref{sec:solucao} we review a non trivial solution to the problem of how the model goes into the SM for the known particles at tree level. The last section is devoted to our conclusions. 

\section{The models}
\label{sec:themodels}

\subsection{From 3-2-1 to 3-3-1}
\label{subsec:321331}
The 3-3-1 models are interesting because they provide a partial answer to the problem of the origin of three generations~\cite{Singer:1980sw,Pisano:1991ee,Foot:1992rh,Frampton:1992wt}: these models are free of anomalies only if the number of generation is three or multiple of three. This solution to the problem of the number of families is not definitive since it has been changed to: who orders these representation contents? However, the fact that 3-3-1 models allow us to predict that number has to be appreciated. After all, most models do not have this property. 

We think that these models deserve more attention since they include particles that appear also in other extensions of the SM, for instance as we stressed above: several scalar fields in doublets and triplets of $SU(2)\otimes U(1)_Y$, new neutral and charged vector fields, and last, but not least, new vector-like fermions (in some cases with exotic electric charge) that are singlets of the SM symmetries.

In these models the defining sector is the leptonic one. After having chosen this sector, quarks are adjusted so that the model is free of anomalies. In fact, these models can be considered not just as extensions of the SM, but as alternatives to it. For example, we can say that the degrees of freedom already discovered are those that coincide with those of the SM. In order to appreciate this, let us recall the lepton representation content of the SM where the lepton sector transform under $SU(3)_c\otimes SU(2)_L\otimes U(1)_Y$, as left-handed doublets and right-handed singlets:
\begin{eqnarray}
\left( 
\begin{array}{c}
\nu^\prime_l\\ l^\prime
\end{array}\right)_L\sim(\mathbf{1},\textbf{2},-1),\quad l^\prime_{l_R}=(l^{\prime c}_l)_L\sim(\mathbf{1},\textbf{1},-2). 
\label{smrc}
\end{eqnarray}
where we define $Q=(\tau_3+Y)/2$, being $\tau_3$ the resepctive Pauli matrix and $Y$ the hypercharge. Primed fields denote symmetry eigenstates.

It is interesting that instead of Eq.~(\ref{smrc}),  we can put all the known leptons together in a triplet (the adjoint repsesentation of $SU(2)$ is discarded since in this case the only neutral vector is the photon~\cite{Georgi:1972cj}). For instance in the minimal 3-3-1 model (m331) leptons transform as~\cite{Pisano:1991ee,Foot:1992rh,Frampton:1992wt}:
\begin{equation}
SU(2)_L \to SU(3)_L:\left( 
\begin{array}{c}
\nu_{l^\prime}\\ l^\prime\\ l^{\prime c}
\end{array}\right)_L\sim\textbf{3},
\end{equation}
Does not seem more natural as a model of leptons? After all, if all the leptons  are massless before the spontaneous symmetry breaking, why not put them all together in the same multiplet? This model is called minimal because only the already known leptons are included. Quarks, as we said above, must be adapted in such a way that the model is free of anomalies only if three generations are considered and $SU(3)_L\to SU(3)_L\otimes U(1)_X$. Notice that for leptons the symmetry is just SU(3), since X=0 for them, and electric charge is quantized in the sense that $\textrm{Tr}Q=0$. The $U(1)_X$ factor is needed to include quarks and it seems that the quantization of the electric charge is lost. Notwithstanding, if other conditions are considered it has been shown that electric charge is quantized independently of the nature of neutrino, Dirac or Majorana~\cite{deSousaPires:1998jc}.

In fact, it is interesting to recall that some of the first models of leptons were based on gauge theories with $SU(3)$ and considered the leptons in triplets~\cite{Salam:1964ry} but in the manner of Konopinski and Mahmoud: the leptons are $(\nu_e\,e^-\,\mu^+ )$ while the anti-leptons are $(\nu^c_e\,e^+\,\mu^- )$~\cite{Konopinski:1953gq}.
Even in Ref.~\cite{Salam:1968rm}, most of it is dedicated to a lepton model with $SU(3)$ symmetry, leaving only for the end to consider the model that was known as the Weinberg-Salam-Glashow electroweak model, with the representation content in Eq.~(\ref{smrc})
with the gauge symmetry $SU(2)_L\otimes U(1)_Y$~\cite{Glashow:1961tr,Weinberg:1967tq}. This was the beginning of the now called standard electroweak model which was builded by many people, theoretically and experimentally~\cite{Rosner:2002xi} and which is now a textbook theme.

On the other hand, if right-handed neutrinos do exist, a possible representation content of the SM with this type of neutrinos is 
\begin{eqnarray}
\left( 
\begin{array}{c}
\nu_{l^\prime}\\ l^\prime
\end{array}\right)_L\sim(\mathbf{1},\textbf{2},-1)\quad l^\prime_R\sim (\mathbf{1},\textbf{1},-2),\;\; (\nu_{l^\prime R})^c=(\nu^c_{l^\prime})_L\sim(\mathbf{1},\textbf{1},0),
\end{eqnarray}
 we have the option of the $\nu_R$331 model~\cite{Montero:1992jk,Foot:1994ym}:
\begin{equation}
SU(2)_L\otimes U(1)_Y \to SU(3)_L\otimes U(1)_X:\left( 
\begin{array}{c}
\nu_{l^\prime}\\ l^\prime\\ \nu^c_{l^\prime}
\end{array}\right)_L\sim(\textbf{3},-1/3),\quad l^\prime_R\sim(\textbf{1},-1).
\end{equation}
Again, quarks are accommodated in such a way that the model is anomaly free.
	
There are several possibilities for the representation content in the 3-3-1 models. As we said before, the defining sector is the leptonic one. The  electric charge operator, $Q$, depends mainly on two parameters, $\beta$ and $X$, and, secondly for a given $\beta$ we can obtain a different representation content by choosing different values of $X$.   
Notice that, like the SM, the 3-3-1 models are chiral.

\subsection{The electric charge operator}
\label{subsec:qo}

In principle, the electric charge operator of the models are parametrized by three parameters
$a,\beta$ and $X$:
\begin{eqnarray}
Q_{\beta}( X)= aT_3+\beta T_8+X\mathbf{1},
\label{charge}
\end{eqnarray}
the $\beta$ parameter introduced in Refs.~\cite{Ozer:1996jc,Ochoa:2005ih,Ochoa:2005ch}. Hereafter we use $a=1$, and $\beta=\pm\sqrt{3},\pm\,1/\sqrt3$:
\begin{equation}
Q_{-\sqrt{3}}(X)=\left(
\begin{array}{ccc}
X & 0 & 0 \\
0 & -1+X & 0 \\
0 & 0 & 1+X 
\end{array} 
\right) \;(a),\;
Q_{-\frac{1}{\sqrt{3}}}(X)\!=\!\left(
\begin{array}{ccc}
\frac{1}{3}+X & 0 & 0 \\
0 & -\frac{2}{3}+X & 0 \\
0 & 0 & \frac{1}{3} +X
\end{array} 
\right)\,(b) 
\label{m331331rn}
\end{equation}
and 	
\begin{equation}
Q_{\sqrt{3}}(X)=\left(
\begin{array}{ccc}
1+X & 0 & 0 \\
0 & X & 0 \\
0 & 0 & -1+X 
\end{array} 
\right)\,(a),\;
Q_{\frac{1}{\sqrt{3}}}(X)\!=\!\left(
\begin{array}{ccc}
\frac{2}{3}+X & 0 & 0 \\
0 & -\frac{1}{3}+X & 0 \\
0 & 0 & -\frac{1}{3}+X 
\end{array} 
\right)\,(b) 
\label{331xy}
\end{equation}
With the notation of Eq.~(\ref{charge}), the two most studied models are those with $\beta=-\sqrt{3}$ and $\beta=-1/\sqrt3$. Anyway, the other models~~\cite{Pleitez:1994pu,Ponce:2001jn} can be obtained of the latter ones.

\subsection{3-3-1 models with $\beta=-\sqrt{3}$ and $\beta=-1/\sqrt{3}$}
\label{subsec:2331models}

We consider the full representation content of the 3-3-1 models with $\beta=-\sqrt3$ and $\beta=-1/\sqrt3$ recalling that a model is defined not only by the gauge (and global) symmetries but also by its representation content. Below, all fields are symmetry eigenstates but we omit any special notation for them. 

\subsubsection{Minimal 331 model}
\label{subsubsec:m331}

Below we omit the prime but all the fields are symmetry eigenstates.
Let us consider the representation content of the m331 model with 
$\beta=-\sqrt{3}$ and leptons with $X=0$ ~\cite{Pisano:1991ee,Foot:1992rh,Frampton:1992wt}:

\textbf{Leptons}:
\begin{equation}
	\Psi_{lL}=\left(\begin{array}{c}
		\nu_l\\l^- \\ l^+
	\end{array}\right)_L\sim(\mathbf{1},\mathbf{3},0).
\end{equation}
Here $l=e,\mu,\tau$. $\nu_{lR}\!\! \sim({\bf1},{\bf 1},0)$ can be added if necessary.  

\textbf{Quarks}: 
\begin{equation}
Q_{mL}=\left(\begin{array}{c}
d_m\\ -u_m\\ j_m
\end{array}
\right)_L	\sim({\bf3},{\bf 3}^{*},-1/3),\quad m=1,2;\quad Q_{3L}=\left(\begin{array}{c}
u_3\\ d_3\\ J
\end{array}
\right)_L	\sim({\bf3},{\bf 3},2/3)
\label{quarksm331}
\end{equation}
and right-handed singlets
$u_{\alpha
R}\sim({\bf3},{\bf 1},2/3)$, $d_{\alpha R}\sim({\bf3},{\bf
1},-1/3),\,\alpha=1,2,3$, $j_{mR}\sim({\bf3},{\bf 1},-4/3)$, and
$J_{R}\sim({\bf3},{\bf 1},5/3)$.  Notice that the electric charge of the quarks $j_m$ and $J$ are (in units of $\vert e\vert$) are -4/3 and 5/3, respectively.
		
\textbf{Scalars}:
\begin{equation}
\eta=\left(\begin{array}{c}
\eta^0\\ \eta^{-}_1\\ \eta^+_2
\end{array}\right)\sim({\bf1},{\bf3},0),\;\rho=\left(\begin{array}{c}
\rho^+\\ \rho^0\\ \rho^{++}
\end{array}\right)\sim({\bf1},{\bf3},1),\;\chi=\left(\begin{array}{c}
\chi^-\\ \chi^{--}\\ \chi^0
\end{array}\right)\sim({\bf1},{\bf3},-1),
\end{equation}
and the sextet $\textbf{S}\sim({\bf1},{\bf6},0)$.

This model, as we said before, will be dubbed m331 for short. It is also possible the following representation contents: lepton triplets $(\nu_{l} \, l^- E^+_l)^T_L$, singlets $l^-_R, E^+_{R}$ ($\nu_R$); being $E_{l}$ new charged heavy leptons. In this case the introduction of the sextet in not mandatory. The model is usually called 3-3-1 model with heavy leptons, or 331HL for short~\cite{Pleitez:1992xh,Correia:2015tra}. In both cases the quarks sector is the same.

\subsubsection{Model with right-handed neutrinos in the lepton triplets}
\label{subsubsec:nur331}

In this case $\beta=-1/\sqrt3$, and leptons have $X=-1/3$~\cite{Montero:1992jk,Foot:1994ym}, the minimal representation content is:
	
\textbf{Leptons}:
\begin{equation}
\Psi_{lL}=\left(\begin{array}{c}
\nu_l\\l^- \\ \nu^c_l
\end{array}\right)_L\sim(\mathbf{1},\mathbf{3},-1/3).
\end{equation}
Here $l=e,\mu,\tau$; and $l_{R}\sim({\bf1},{\bf 1},-1))$.

\textbf{Quarks}: 
\begin{equation}
Q_{mL}=\left(\begin{array}{c}
d_m\\ -u_m\\ d^\prime_m
\end{array}
\right)_L	\sim({\bf3},{\bf 3}^{*},-1/3),\quad m=1,2;\quad Q_{3L}=\left(\begin{array}{c}
u_3\\ d_3\\
u^\prime
\end{array}
\right)_L	\sim({\bf3},{\bf 3},2/3),
\label{quarksmnur}
\end{equation}
and right-handed singlets $u_{\alpha
R}\sim({\bf3},{\bf 1},2/3)$, $\alpha=1,2,3,4$, and $d_{\gamma R}\sim({\bf3},{\bf
1},-1/3),\,\gamma=1,2,3,4,5$. 

\textbf{Scalars}: 
\begin{equation}
\eta=\left(\begin{array}{c}
\eta^+_1,\\\eta^0,\\ \eta^+_2
\end{array}\right)\sim({\bf1},{\bf3},2/3),\;
\rho=\left(\begin{array}{c}
\rho^0_1,\\ \rho^-,\\ \rho^0_2
\end{array}\right)\sim({\bf1},{\bf3},-1/3),
\end{equation}
A third triplet $\chi=(\chi^0\,\chi^-\,\chi^0)^T\sim({\bf1},{\bf3},-1/3)$ may be included to avoid the flavor changing neutral currents in the Higgs sector.
A different model is obtained if the lepton triplets are $(\nu_{l}\,l\,N)_L$ and singlets $l_R$, $\nu_R$ and $N_R$~\cite{Singer:1980sw}. This model is usually called 3-3-1 model with right-handed neutrinos, but $N_L$ may be considered a lepton not an anti-lepton. 


Next, we will show that models with the same values of $\beta$ and $X$, but different leptonic representation content, have different phenomenology in this sector.

\subsection{Same $\beta$ and $X$ and different phenomemnology}
\label{subsec:dbeta}

Models with the same value of $\beta$ and $X$, have different phenomenology in the lepton sector because they may have different representation content in the lepton and scalar sectors. For instance if $\beta=-\sqrt{3}$ and $X=0$ there are two possibilities:
\begin{eqnarray}
\left( 
\begin{array}{c}
\nu_l\\ l^-\\ l^+
\end{array}\right)_L \;\;(a),\quad \left( 
\begin{array}{c}
\nu_l\\ l\\ E^+_l
\end{array}\right)_L\;\;(b),
\end{eqnarray}
model (a) does not need any singlets $\nu_{lR}$ but they can be included if necessary; (b) needs the singlets $l^-_R$ and $E^+_{lR}$ and right-handed neutrinos may also be added if necessary. Notice also that $E^+_l$ may be considered a \textit{lepton}, then lepton number is conserved in model (b) but not in (a). Similar to the Konopinski and Mahmoud scheme~\cite{Konopinski:1953gq}: leptons are $\nu,\,l^-$ and $\mu^+\to E^+_l$; anti-leptons are  $\nu^c,\,l^+$ and $\mu^-\to E^-_l$. This point was not considered in Ref.~\cite{Fonseca:2016xsy}. The minimal scalar sector are different also: Model (a) needs besides the three triplets, a scalar sextet, while model (b) only three triplets.

Similarly, two models with $\beta=-1/\sqrt3$ and $X=-1/3$ are different from the phenomenological point of view in the lepton and in the scaralr sectors: 
\begin{eqnarray}
\left( 
\begin{array}{c}
\nu_l\\ l\\ \nu^c_l
\end{array}\right)_L \;\;(a),\quad \left( 
\begin{array}{c}
\nu_l\\ l\\ N_l
\end{array}\right)_L\;\;(b),
\end{eqnarray}
model (a) needs only the singlet $l^-_R$ (each one per generation) and a scalar sextet; model (b) needs the singlets $l^-_R,N_{lR}$ (per generation) and $\nu_{lR}$ may be introduced if necessary. Again, $N_l$ may be a lepton.	 
For instance, effects in $(g-2)_\mu$ for the same value of $\beta$ but different particle contents see~\cite{deJesus:2020ngn}. Model (a) needs a sextet to give mass to neutrinos, model (b) needs only triplets.

\subsection{Some predictions of the models}
\label{subsec:predictions}

The minimal versions of these models have interesting phenomenological consequences. For example 
\begin{enumerate}
\item[(a)] Explain the number of generations (multiple of three or just three if asymptotic freedom is valid for the exotic quarks) \cite{Pisano:1991ee,Singer:1980sw}.
\item[(b)] Explain the quantization of the electric charge~\cite{deSousaPires:1998jc}, for 3-4-1 models see~\cite{Cabarcas:2013hwa}; and Ref.~\cite{Palcu:2021jst} for	$3\!\!-\!\!n\!\!-\!\!1$ models.
	
\item[(c)] They incorporate (almost) naturally the Peccei-Quinn symmetry~\cite{Pal:1994ba,Dias:2003zt,Dias:2004hy,Dias:2002hz}.
\item[(d)] Models with $\beta=-\sqrt{3}$~\cite{Pisano:1991ee,Foot:1992rh,Frampton:1992wt} predicts quarks with exotic electric charge $F^Q=J^{5/3}, j^{-4/3}$ and doubly charged vector boson, $U^{++}_\mu$, carrying lepton number $L=-2$~\cite{Pleitez:1993gc}, and models with $\beta=-1/\sqrt3$ predict new $u$- and $d$-type quarks and a neutral vector boson $\mathcal{Z}^0$ carrying also two units of lepton number~\cite{Singer:1980sw,Montero:1992jk,Foot:1994ym}. 
	
\item[(e)] If we assume that the condition 
\begin{equation}
\frac{g_X}{g_{3L}}=\frac{s^2_W}{1+\frac{4}{\sqrt{3} }\beta s^2_W}
\label{condition}
\end{equation}
is valid for any energy, then, the models with $\beta=-\sqrt{3}$ explain why $\sin^2\theta_W<1/4$~\cite{Frampton:1992wt,Ng:1992st} which occurs at energies of the order of the TeVs~\cite{Dias:2004dc} or $\sin^2\theta_W>1/4$~\cite{barela}.
For models with $\beta=-1/\sqrt3$ the condition is when $\sin^2\theta_W$ is lower or greater than $3/4$ which occurs at a rather high energy. 
	
\item[(f)] They incorporate many of the multi-Higgs extensions of the standard model, singlets (neutral or charged), several doublets, and triplets~\cite{DeConto:2015eia}.
\item[(g)] All these models accommodate new sources of $C\!P$ violation~\cite{Promberger:2007py,Buras:2015kwd,Buras:2012dp}. 
\item[(h)] They have a neutral real vector boson $Z^\prime$ which has interesting contribution to flavor physics~\cite{Buras:2014yna,Buras:2016dxz}.
\item[(i)] The model has candidates for dark matter~\cite{Dong:2015rka,Dong:2015dxw}. For instance, if active neutrinos are pure Majorana particles, the right-handed neutrinos are not necessary in the m331 and they may be introduced just as dark matter candidates~\cite{Canetti:2012vf}.
	
\item[(j)] These models allow partial dynamical symmetry breaking~\cite{Das:1999hn,Doff:2015ukb}.
	
\item[(k)] In some versions of the models there are new source of $\vert \Delta L\vert=2$ processes~\cite{Fonseca:2016xsy}. In the m331 model these effects occur even without the scalar sextet in the model. For instance there are new contribution to the $(\beta\beta)_{0\nu}$ decay~\cite{Pleitez:1993gc}.
\end{enumerate}

\section{The standard model projection}
\label{sec:smp}

\begin{center}
\begin{figure}[!h]
\includegraphics[width=20cm]{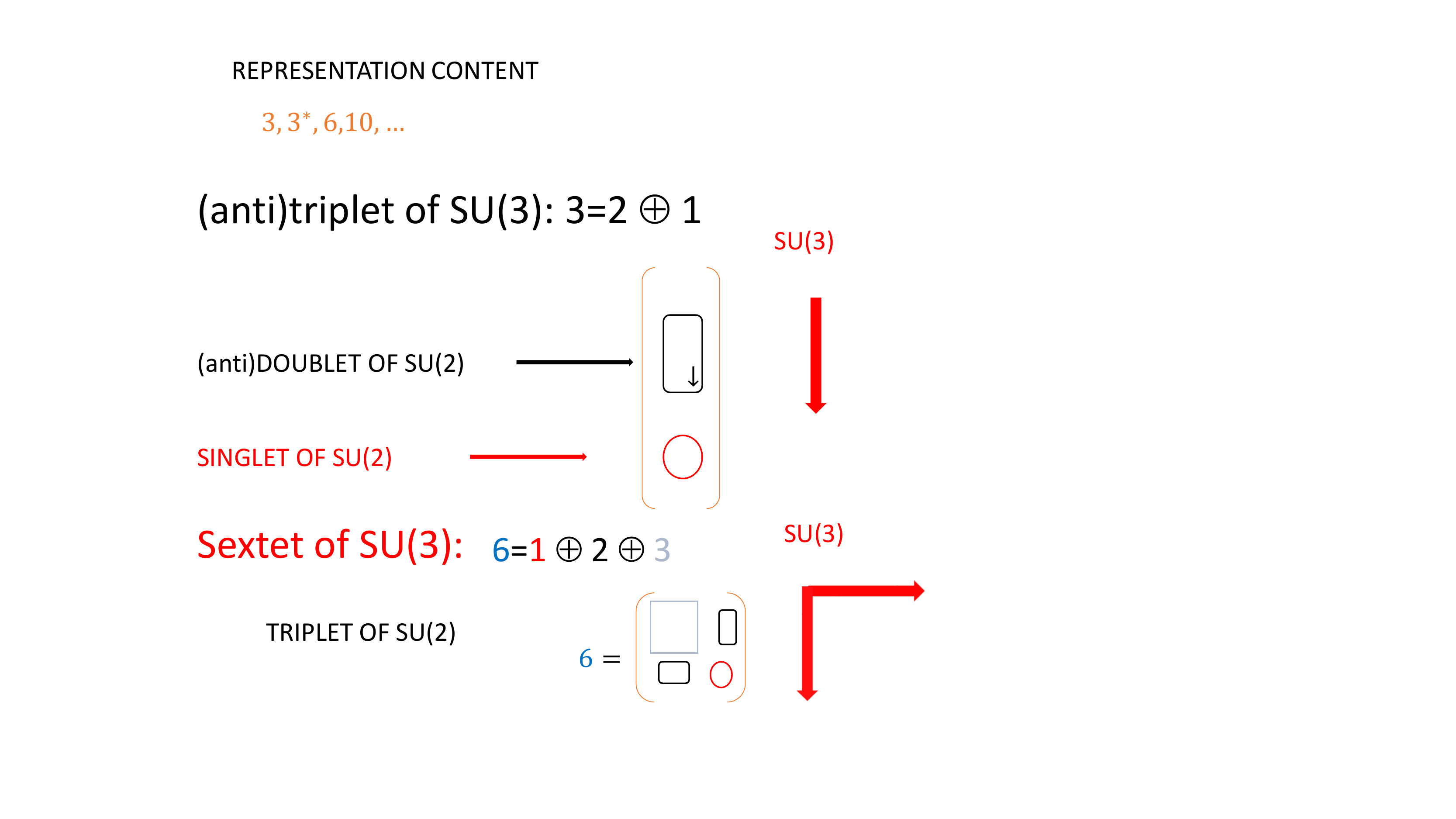}
\caption{Projection of the representation content of 331 models into the SM symmetries. }
\label{fig:smprojection} 
\end{figure}
\end{center}

Although the 3-3-1 model contain the complete representation of the SM, six quarks and six leptons, the $W^\pm$ and $Z^0$ bosons, their projection in the symmetry of the SM also includes, as it was emphasized above, multiplets that, although they are not in the minimal version of SM, they have been considered in some of its more motivated extensions: several Higgs doublets; fields singlets of $SU(2)$: neutral and charged scalars, and also new quarks and leptons. Moreover, doublets of vector and scalar fields carrying lepton $L$ number and exotic electric charge.
In terms of the  $SU(2)_L\otimes U(1)_Y$ symmetries, the representation contents of any 3-3-1 model are as is shown in Figure~\ref{fig:smprojection}.

In this section we consider the projection of the representation content of the m331 model ($\beta=-\sqrt3,\;X=0$) on $SU(3)_c\otimes SU(2)_L\otimes U(1)_Y$. 
Firstly, let us consider the SM projection of the already known quarks and leptons. We have to consider the three generations (recall that the models are anomalous with only one or two generations).

Leptons are as in Eq.~(\ref{smrc}) while quarks transform as follows
\begin{eqnarray}
Q_L=
\left(
\begin{array}{l}
u_i\\ d_i
\end{array}
\right)_L\sim(\mathbf{3},\mathbf{2},1/3),\;\; u_{iR}\sim (\mathbf{3},\mathbf{2},4/3),\; d_{iR}\sim (\mathbf{3},\mathbf{1},-2/3),
\end{eqnarray}
$i=1,2,3$.
The models have also vector-like quarks singlets of $SU(2)_L$.
In particular, the m331 model have vector-like quarks
\begin{eqnarray}
J_{L,R}\sim (\mathbf{3},\mathbf{1},10/3),\quad j_{m(L,R)}\sim  (\mathbf{3},\mathbf{1},-8/3),\;m=1,2,
\end{eqnarray}
with electric charge $+5/3$ and $-4/3$, respectively. 

The scalar sector has the following multiplets transforming under the SM symmetries. Scalar doublets:
\begin{eqnarray}
\left(\begin{array}{c}
\eta^0\\ -\eta^-
\end{array} 
\right),\;\left(\begin{array}{c}
\rho^+\\ \rho^0
\end{array} 
\right),\;\left(\begin{array}{c}
s^+_2\\ s^0_2
\end{array} 
\right),
\end{eqnarray}
with $Y=-1,+1,+1$, respectively. A scalar triplet $T$ of $SU(2)$ with $Y=2$,
\begin{equation}
\mathbf{T}=\vec{\tau}\cdot\vec{t}=\left( \begin{array}{cc}
T^+ & \sqrt{2}T^{++} \\
\sqrt{2}T^0 & -T^+ 
\end{array}
\right),
\label{triplet}
\end{equation}
scalar singlets $\eta^+_2,\rho^{++},\chi^0,S^{--}_2$ with $Y=+2,+4,0,+4$, respectively.  Moreover, there are two doublets with $Y=+3$ of vector and scalar fields, $\mathbf{V}_\mu$, and  $\mathbf{\Xi}$, respectively:
\begin{equation} 
\mathbf{V}_\mu=\left(
\begin{array}{c}
 U^{++}_\mu \\ V^+_\mu \\
\end{array}
\right), \quad \;
\mathbf{\Xi}=\left(\begin{array}{c}
\chi^{++}\\ \chi^+
\end{array}	
\right).
\label{nova}
\end{equation}
The vector doublet $\mathbf{V}_\mu$ is in an octet of $SU(3)$ of gauge bosons but, after the breakdown of symmetry they become matter fields. In general, the doubly charged scalar appear as member of a triplet $Y=2$.
Finally, there is a neutral boson $Z^\prime_\mu$ which gain mass by the VEV of the singlet $\chi^0$.

The interactions involving the vector bosons in Eq.~(\ref{nova}) with the known quarks are
\begin{equation}
\mathcal{L}_{Vj(J)}=\overline{Q}_{iL}K_{im}\gamma^\mu j_{mR}\mathbf{V}_{\mu}+\overline{Q}_{iL}\gamma^\mu K_iJ_R \mathbf{\tilde{V}}_{\mu}+H.c.,
\label{oba1}
\end{equation}
where $i=1,2,3$, $m=1,2$ and we have defined  $\mathbf{\tilde{V}_\mu}=i\tau_2\mathbf{V}^*_\mu$.
As these interactions are inspired in the m331 model, we define $K_{im}=(g/\sqrt{2})\delta_{im}$ and $k_i=g/\sqrt{2}$. In the mass eigenstates basis (unless a factor of $g/\sqrt2$) we write the interactions above as
\begin{eqnarray}
\mathcal{L}_{Vj(J)}&=&\bar{u}_{iL}(V^{u\dagger}_LO)_{ik}\gamma^\mu j_{kR}
U^{++}_\mu-\bar{d}_{iL}(V^{d\dagger}_LO)_{ik}j_{kR}V^+_\mu\nonumber \\&+&
\bar{u}_{iL}(V^{u\dagger}_L)_{iJ}\gamma^\mu J_LV^{-}_\mu-\bar{d}_{iL}(V^{d\dagger}_L)_{iJ}\gamma^\mu J^c_L U^{--}_\mu+H.c.
\label{oba2}
\end{eqnarray}
$i=1,2,3;\,k=1,2$ and the index $J$ is fixed.

Notice that these vector-like quarks have transitions to single charged vector bosons as 
\begin{eqnarray}
	b\to j+V^+_\mu,\quad  t\to J+V^-_\mu,
\end{eqnarray}
i.e., quarks $j$ and $J$ couples only with quarks $d(u)$ thorughout $V^+_\mu$ not with $W^+_\mu$; and the transitions to doubly charged vector bosons are
\begin{eqnarray}
	t\to j+U^{++}_\mu,\quad b\to J+U^{--}_\mu.
\end{eqnarray}

The vector doublet $\mathbf{V}_\mu$ couples also with leptons (unless a  factor of $g/\sqrt2$):
\begin{eqnarray}
\mathcal{L}&=&\bar{L}\gamma^\mu l^c_L\mathbf{V}_\mu+H.c.\nonumber \\&=&
\bar{\nu}_{iL}\gamma^\mu(V^\nu_LV^{l*}_R)_{ij}l^c_{jL}V^-_\mu-\bar{l}_{iL}\gamma^\mu (V^{l\dagger}_LV^{l*}_R)_{ij}l^c_{jL} U^{--}_\mu+H.c.,
\label{obal}
\end{eqnarray}
where in the second line we have written the interactions in the mass basis.

Notice the transitions like 
\begin{equation}
\nu_L\to l^+_L+V^-_\mu,\quad l^-_L\to l^+_L+U^{--}_\mu.
\label{obal2}
\end{equation}

The scalar $\mathbf{\Xi}$ couples only with quarks as follows:
\begin{equation}
-\mathcal{L}_{\Xi j(J)}=\bar{Q}_{nL}G^j_{nm} j_{mR}\epsilon_{kl}\mathbf{\Xi}^*_{l}+\bar{Q}_{kL} G^J_k J_R\mathbf{\Xi}+H.c.,
\label{oba3}
\end{equation}
or, in terms of the mass eigenstates
\begin{eqnarray}
-\mathcal{L}_{\Xi j(J)}&=&\bar{u}_{iL}(V^{u\dagger}_LG^jO)_{ij}j_{jR}\chi^{++}-
\bar{d}_{iL}(V^{d\dagger}_LG^jO)_{ij}j_{jR}\chi^+\nonumber\\&+&
\bar{u}_{iL}(V^{u\dagger}G^J)_iJ_R\chi^-+\bar{d}_{iL}(V^{d\dagger}_LG^J)_iJ_R\chi^{--}+H.c.,
\label{oba4}
\end{eqnarray}
and we have transitions as 
\begin{equation}
t_L\to j_R+\chi^{++},\quad b\_Lto J_R+\chi^{--},
\label{oba5}
\end{equation}
and
\begin{equation}
t_L\to J_R+\chi^-,\quad b_L\to j_R+\chi^+.
\label{oba6}
\end{equation}

Since all 3-3-1 models have a real neutral vector boson $Z^{\prime0}_\mu$ and a complex neutral scalar $\chi^0$ we have 
\begin{equation}
\mathcal{L}_Z=(\partial_\mu+i\xi_\chi Z^{\prime0})\chi^{0*}(\partial_\mu-i\xi_\chi Z^{\prime0})\chi^0,
\label{zp}
\end{equation}
The neutral vector field cannot has a kinetic mixing with $A$ or $Z$ since these term does not exist in the m331. Hence there is a parity-like conservation. Under this operation $Z^{\prime0},\chi^0$ and exotic fields are odd while the field that also belong to the SM spectra are even.   

The scalar potential includes besides the doublets with $\vert Y\vert=1$, the complex triplet and also the singlets $\eta^-_2,\rho^{++},\chi^0$, and the doublet $\mathbf{\Xi}$.

Moreover, it was shown in Ref.~\cite{Pleitez:1993gc} that the model is invariant under local conservation of $B+L$.
The new fields carry an extra quantum number related to $X=B+L$: 
\begin{eqnarray}
X(J)=X(j_m)=+\frac{7}{3},\quad
X(V^-,U^{--},\chi^-,\chi^{--},\eta^+_2)=+2, \quad X(H^{--}_2,\rho^{++},T)=-2,
\label{lnumber}
\end{eqnarray}
and the particles belonging to the SM carry their usual $B$ and $L$ charges.

Summaryzing, the SM projection of the 3-3-1 models are into
\begin{equation}
SU(2)_L\otimes U(1)_Y\rtimes U(1)_{B+L}.	
\label{upa}
\end{equation}
Notice that the $U(1)_{B+L}$ factor is a semidirect product.

The above interactions are considered in of the context of the symmetries in Eq.~(\ref{upa}). Notice that all the extra fields, but the doublets $(U^{++}_\mu\, V^+_\mu)$, have been already considered as extensions of the SM. For instance
\begin{itemize} 
\item The SM plus three scalar doublets with $Y=+1$~\cite{Machado:2010uc}.
\item A singly charged singlet scalar with $Y=4$, $\eta^+_2$, inducing Majorana masses for neutrino at 1-loop level. Under certain conditions this mechanism can be implemented in the m331 model~\cite{Zee:1980ai,Machado:2017flo}.
\item A doubly charged singlet $S^{--}_1$ has been considered  for generating Majorana neutrino masses at the 2-loop level~\cite{Zee:1985id,Babu:1988ki}.
\item A complex triplet $T$ generating the type-II see saw model~\cite{Schechter:1980gr,Cheng:1980qt}.
\item The model with $\beta=-\sqrt{3}$ and $X=0$ but with heavy charged leptons that transform as singlet under $SU(2)_L\otimes U(1)_Y$, $E_{lL},E_{lR}$~\cite{Pleitez:1992xh} could implement the see-saw mechanism for lepton masses and explain the muon $(g − 2)$~\cite{Lee:2021gnw}.
\item The scalar doublet $\mathbf{\Xi}$ in Eq.~(\ref{nova}), which is part of a triplet of $SU(3)$, has also been considered as extension of the SM~\cite{Aoki:2011yk}, for instance, in a new scotogenic model~\cite{Puerta:2021hfl}. 
\end{itemize}

Since all these extensions of the SM are embedded in the m331, the fact that these models have many scalar fields may be not a problem, but an interesting feature. 

In models with $\beta=-1/\sqrt{3}$ the standard model projection implies singlet fermions: neutral $N_L$, or $(\nu_L)^c$, and quarks $u^\prime_4$ and $d^\prime_4,d^\prime_5$ (symmetry eigenstates). In the these models it is possible to have mixing among all quarks of the same charge or, the mixing can be avoided if discrete symmetries are introduced. We stress that as remarked in Sec.~\ref{subsec:dbeta}: a model with $(\nu^c_l)_L$ in the triplet is a different model from that with one with  neutral fermion, $N_{lL}$, in each triplet.

At this point, we can ask ourselves, what defines a 3-3-1 model? 
Of course, the 331 symmetry is necessary but not sufficient, as we already said, a model is characterized not just by its symmetries but also by its particle content. 
Firstly, they are chiral models, secondly, the triangle anomalies are canceled out only with a number of generations multiple of 3, or equal to 3 if the additional quarks contribute to the running of $\alpha_{QCD}$; Finally, 
they have a projection into the SM which includes besides the particles of the SM, some of its more motivated extensions, as in shown in Fig.~\ref{fig:smprojection}. These requirements are what define the 3-3-1 models. Moreover, the fact that the projection into the 3-2-1 symmetry include the more studied extensions of the SM gives to them a rationale.

\section{old electroweak models with $SU(3)\otimes U(1)$ symmetries }
\label{sec:interlude}

It is good at this point to make an interlude to comment on some electroweak models with $SU(3)\otimes U(1)$ symmetries proposed in the early 1970s. These models are not of the type now called 3-3-1 models although they have the same electroweak symmetries and the charge operator defined as in Eq.~(\ref{charge}) they are anomalous or are not chiral and,  last but not least, have not the projection into the SM with the three known generations.
These models have $\beta=+1/\sqrt{3}$, with the electric charge operator given in Eq.~(\ref{331xy})(b) but different values of $X$. 
\begin{itemize}
\item Two models, one with $SU(3)\otimes U(1)$ and the other with $SU(3)\otimes SU(3)\otimes U(1)\otimes U(1)$ was proposed in Ref.~\cite{Schechter:1974dy}. The model has $X=-2/3$ for leptons but only two generations of leptons (including two heavy leptons) are considered, the left-handed fields are in triplets, and the right-handed components of the charged fields in singlets. In the quark sector the known left-handed quarks $u,d,s$  have $X=0$ and are in triplet; the respective right-handed components are in singlets. Obviusly the model is anomalous. In the chiral $SU(3)\otimes SU(3)$ model both left-handed and right-handed are in triplets of each $SU(3)$ group. The representation content of this chiral model is the same as in the first model. For this reason the model is also anomalous. In fact, that the model should be renormalizable, modulo anomalies, was recognized in Ref.~\cite{Schechter:1974mq}.
Besides, these models do not have the projection into the SM as in Fig.~\ref{fig:smprojection} and Sec.~\ref{sec:smp}. 
 
\item In Ref.~\cite{Lee:1977qs} it was proposed an electroweak model with $SU(3)\otimes U(1)$ gauge symmetry, and with the values of $\beta$ and $X$ for leptons as in the first model of Ref.~\cite{Schechter:1974dy}, but now three lepton and two quark generations are consider; the leptons of each chirality are in triplets with $X=-2/3$ and there are also left-handed neutral singlets. The left- and right- handed quarks $u,d,s$ are in triplets. There are extra heavy fermions which are odd under a $R$ symmetry. Although the model is free of anomalies since the model is not quiral i.e, is a vector model, it does not have the same $SU(2)\otimes U(1)$ representation content, for instance right-handed electron belongs to a triplet together with two other heavy leptons. Similarly with the right-handed muon. The scalar sector consists of one triplet with $X=-2/3$, and one octet with $X=0$. A similar model but strictly vector-like triplets was proposed in Ref.~\cite{Segre:1976rc}. 

\item The $SU(3)\otimes U(1)$ of Ref.~\cite{Langacker:1977ae} considered leptons in anti-triplets with $X=-1/3$ and quarks in triplets with $X=0$. In this model the left- and right-handed electron, and also the muons, are in anti-triplets; but the left-handed tau lepton is also in a triplet but its right-handed components are in singlet. Although the model is anomaly free (because of its vector-like structure) it does not has the SM projection in Sec.~\ref{sec:smp}. 

\item The model of Ref.~\cite{Lee:1977tx} has the following representation content: a quark family consists of one triplet with $X=0$ and one singlet of each chirality. A lepton family consists of one 
triplet of each chirality, and one $X=0$ left-handed
singlet. The model is free of nomalies due to the quasi-vectorlike character but again it has not the projection into the SM representation as in Fig.~\ref{fig:smprojection}.
\end{itemize}

The motivation of these models was the possibility that the events with three muon do exist (see below).
We see that the old models with $SU(3)\otimes U(1)$ symmetry cannot be called 3-3-1 models in the sense defined at the end of Sec.~\ref{sec:smp}: either they were anomalous as in Ref.~\cite{Schechter:1974dy}, or they are anomaly free because of the vector-like character. In fact, the first model with simetry $SU(3)_L\otimes U(1)_X$ and $\beta=-1/\sqrt3$, in which anomalies were cancel out because the number of triplets is equal to the number of antitriplets was proposed in Ref.~\cite{Georgi:1978bv}. However, although the model has six quarks, they are two with electric charge 2/3, and four -$1/3$. That is, the model consider only two of the known generations: $(u,d,s,c)$ and $(\nu_e,e,\nu_\mu,\mu)$ plus exotic fermions. In order to cancel the anomalies is necessary add unkown heavy leptons and the antileptons in $SU(2)$ doublets. 
Hence the model has not the SM projection as in Sec.~\ref{sec:smp}.
Moreover, in this model if we add the known third family of quarks, it must be vector-like under the SM group in order not to spoil the anomalies cancelation~\cite{Pleitez:1994pu}. In any case the models do not have the projection into the known standard model with three chiral generations. 

We stress that, the so called 3-3-1 models are chiral models and the anomalies are canceled out among the three generations with the number of $\textbf{3}$ being equal to the number of $\textbf{3}^*$ and apropriately chosen $X$ charges and also, the projections into SM the known particles must be as it was shown in Sec.~\ref{sec:smp}.

\section{Some Challenges}
\label{sec:challenges}

It is well known that models beyond the SM (BSM) have a rich particle spectrum and, in general, they have new interactions which depend on many parameters. For instance, mixing matrices with new $C\!P$ violating phases, Yukawa couplings and the mass of the extra degrees of freedom. The 3-3-1 models are not exception of this, they have many new parameters and fields, but that also makes them phenomenologically robust. Let us remember that the SM would have already been discarded if the Cabibbo angle had not been proposed first, and the CKM matrix later. The radiative corrections were also very important to make the SM consistent with the experimental data, and they will probably also be important in physics BSM when the right model (if any) be discovered.

Going back to the 3-3-1 models, here we will concentrate in some issues that although most of them have already been considered in literature, we think that the last word has not been given yet. Probably, one day it may be necessary to reconsider them. They are
\begin{itemize}
\item The search of exotic  quarks or leptons
\item Exotic baryons
\item Neutrino masses 
\item The old processes that were not confirmed
\item Confusion among neutrinos and antineutrinos
\end{itemize}
Before considering in more detail the challenges in the quark and lepton sectors, we discuss the origin of the parameters of the m331 model.

\section{Parameters of the m331 model}
\label{sec:parameters}

In order to understand why there are many parameters in this sort of models, we summarize the quark and lepton in the context of the m331 model. The model with $\beta=1/\sqrt{3}$ is also briefly considered. 

The quark sector is considered in Subec.~\ref{subsec:quarks} and leptons in Subec.~\ref{subsec:leptons}.

\subsection{Quarks}
\label{subsec:quarks}

In all 3-3-1 models, since two quark generations transform different from the third one, the model needs several scalar triplets.
In the m331 model, the triplets $\eta,\rho$, and $\chi$ are needed in order to give appropriate mass to all quarks and the sextet $S\sim(\textbf{6},0)$ coupling only with leptons, and the triplet $\eta$, allow to leptons to acquiere masses. All of them are given in Subsec.~\ref{subsubsec:m331}. 

The Yukawa interactions in that sector are given by
\begin{eqnarray}
-\mathcal{L}_{Yq}&=&\overline{Q_{mL}}\,[G_{m\alpha}U_{\alpha R}\rho^*+\tilde{G}_{m\alpha}D_{\alpha R}\eta^*]+
\overline{Q_{3L}}\,[F_{3\alpha}U_{\alpha R}\eta+\tilde{F}_{3\alpha}D_{\alpha R}\rho]
\nonumber \\&+&
\overline{Q_{mL}}\,G^j_{mn}j_{nR}\chi^*+
g_J\,\overline{Q_{3L}}\,J_R\chi +H.c.
\label{yukawam331q}
\end{eqnarray}
The mass matrices for the quarks $U$ and $D$ arise when $\rho\to v_\rho/\sqrt{2},\eta\to v_\eta/\sqrt{2}$ and $\chi\to v_\chi/\sqrt{2}$ (we consider all VEVs real) and both have contributions of two triplets $\eta$ and $\rho$. Hence there are FCNC mediated by neutral scalars in the quark sector~\cite{Glashow:1976nt}. In Eq.~(\ref{yukawam331q}) all states are symmetry eigenstates. Since there are two quarks with $Q=(-4/3)\vert e \vert$, they mix by a Cabibbo-like $2\times2$ matrix.  Only the quark $J$ does not mix because it is the only one quark with $Q=(5/3)\vert e \vert$. 

\subsubsection{Quark-neutral scalar interactions}
\label{subsubsec:qnsi}

Firstly, we observe that from Eq.~(\ref{yukawam331q}) we obtain the following mass matrices
\begin{eqnarray}
M^U=\left(\begin{array}{ccc}
rG_{11}&rG_{12}&rG_{13}\\
rG_{21}&rG_{22}&rG_{23}\\
F_{31}&F_{32}&F_{33}
\end{array}
\right)\vert v_\eta\vert,\;\;
M^D=\left(\begin{array}{ccc}
r^{-1}\tilde{G}_{11}&r^{-1}\tilde{G}_{12}&r^{-1}\tilde{G}_{13}\\
r^{-1}\tilde{G}_{21}&r^{-1}\tilde{G}_{22}&r^{-1}\tilde{G}_{23}\\
\tilde{F}_{31}&\tilde{F}_{32}&\tilde{F}_{33}
\end{array}
\right)\vert v_\rho\vert.
\label{massud}
\end{eqnarray}
where $r=\vert v_\rho\vert/\vert v_\eta\vert$. We see that, unless an appropriate symmetry is introduced, say to have the $u$-type quarks in the diagonal basis from the very start, we proceed as usual: 
Denote $U^\prime_{L,R},D^\prime_{L,R}$: the eigenstates of the symmetry and
$U_{L,R},D_{L,R}$ the mass eigenstates. For instance, $U^\prime_L=(u^\prime\, c^\prime\,t^\prime)$ and so on. These fields are related to each other
by the unitary matrices $U_{L,R}=V^U_{L,R}U^\prime_{L,R}$ and $D_{L,R}=V^D_{L,R}D^\prime_{L,R}$
and they diagonalized the respective mass matrices:
\begin{eqnarray} 
\mathbf{V^{D\dagger}_L}M^D\mathbf{V^D_R}=\hat{M}^D,\quad
\mathbf{V^{U\dagger}_L}M^U\mathbf{V^U_R}=\hat{M}^U,\quad
\Large{\mathbf{V_{CKM}=V_L^{D\dagger}}}\mathbf{V_L^U}, 
\label{vdvup1}
\end{eqnarray}
with 
\begin{equation}
\hat{M}^U=\textrm{diag}(m_u,m_c,m_t),\quad \hat{M}^D=\textrm{diag}(m_d,m_s,m_b).
\label{mumd}
\end{equation}

Let us recall how the CKM matrix arises in the context of the SM.
In this model the so called \textbf{GIM} mechanism is automatically implemented: there is no FCNC neither with the Higgs, $H^0$, nor with the vector $Z^0$. For this reason
the $V^{U,D}_L$ matrices survive only through the combination in the CKM matrix, $\mathbf{V_{CKM}=V^{D\dagger}_LV^{U}_L}$. Usually $V^U_L=\mathbf{1}$ is used. Moreover, the matrices $V^{U,D}_R$ completely disappear from the Lagrangian. In fact, in the SM we can use the mass matrices of the quarks being diagonal from the beginning and place the CKM matrix by hand.

However, in the m331 model the matrices $\mathbf{V^Q_L}$ and $\mathbf{V^Q_R}$, $\mathbf{Q}=U,D$, survive in combinations that are different from those  appearing in the definition of the CKM matrix. For instance, in the quark-neutral scalar interactions in the quarks mass eigenstates basis are written as
\begin{equation}
-\mathcal{L}^{nc}_{Yq}=\overline{U}_L\mathbf{K}^U U_R
+\overline{D}_L\mathbf{K}^D
D_R+H.c. ,
\label{yukan}
\end{equation}
with $\mathbf{K}^U= \mathbf{V^U_L}\mathcal{Z}^U\mathbf{V^{U\dagger}_R}$ and $\mathbf{K}^D= \mathbf{V^D_L}\mathcal{Z}^D\mathbf{V^{D\dagger}_R}$,
and $\mathcal{Z}^{U,D}$ are $3\times3$ matrizes involving the Yukawa couplings $G$ and $F$ in Eq.~(\ref{yukawam331q}) and are given by~\cite{Machado:2013jca},
\begin{eqnarray}
\mathcal{Z}^U=\left(\begin{array}{ccc}
G_{11}\rho^0 &G_{12}\rho^0&G_{13}\rho^0\\
G_{21}\rho^0&G_{22}\rho^0&G_{23}\rho^0\\
F_{31}\eta^0&F_{32}\eta^0&F_{33}\eta^0
\end{array}
\right),\;\;
\mathcal{Z}^D=\left(\begin{array}{ccc}
\tilde{G}_{11}\eta^0&\tilde{G}_{12}\eta^0&\tilde{G}_{13}\eta^0\\
\tilde{G}_{21}\eta^0&\tilde{G}_{22}\eta^0&\tilde{G}_{23}\eta^0\\
\tilde{F}_{31}\rho^0&\tilde{F}_{32}\rho^0&\tilde{F}_{33}\rho^0
\end{array}
\right).
\label{yukanma}
\end{eqnarray}
The mass terms in Eq.~(\ref{massud}) appears from Eq.~(\ref{yukanma}), when $\sqrt{2}x^0=v_x+\textrm{Re}\,x^0+i\textrm{Im}\,x^0$, with $x^0=\eta^0,\rho^0$, with $\textrm{Re}\,x^0$ and $\textrm{Im}\,x^0$ are still symmetry eigenstates. For instance, $\textrm{Re}\,\eta^0=\sum_iU_{\eta i}h^0_i,\;\textrm{Re}\rho^0=\sum_iU_{\rho i}h^0_i$ with $h^0_i$ being mass eigenstates with mass $m_i$.
It means that FCNC mediated by neutral scalars is a prediction  in the minimal version of the model, at most they may be controlled not avoided.
			
\subsubsection{Quark-charged scalar interactions}
\label{subsubsec:qsci}

The quark-charged scalar interactions in this model are	
\begin{equation}
-\mathcal{L}^{cc}_{Yq}=\overline{D}_L\mathcal{K}^{DU} U_R
+\overline{U}_L\mathcal{K}^{UD}
D_R+H.c. ,
\label{cc1}
\end{equation}
where $\mathcal{K}^{DU}= \mathbf{V^U_L}\mathcal{Z}^{DU}\mathbf{V^{U\dagger}_R}$ and $\mathcal{K}^{UD}= \mathbf{V^D_L}\mathcal{Z}^{UD}\mathbf{V^{D\dagger}_R}$
with~\cite{Machado:2013jca}
\begin{eqnarray}
\mathcal{Z}^{DU}=\left(\begin{array}{ccc}
G_{11}\rho^- &G_{12}\rho^-&G_{13}\rho^-\\
G_{21}\rho^-&G_{22}\rho^-&G_{23}\rho^-\\
\tilde{F}_{31}\eta^-_1&\tilde{F}_{32}\eta^-_1&\tilde{F}_{33}\eta^-_1
\end{array}
\right),\quad
\mathcal{Z}^{UD}=\left(\begin{array}{ccc}
\tilde{G}_{11}\eta^+_1&\tilde{G}_{12}\eta^+_1&\tilde{G}_{13}\eta^+_1\\
\tilde{G}_{21}\eta^+_1&\tilde{G}_{22}\eta^+_1&\tilde{G}_{23}\eta^+_1\\
F_{31}\rho^+&F_{32}\rho^+&F_{33}\rho^+
\end{array}
\right).
\label{yukanma2}
\end{eqnarray}
Notice that in interactions in Eqs.~(\ref{yukan}) and (\ref{cc1}) the matrices $V^{U,D}_{L,R}$ survive separately.
	
\subsubsection{Quark-vector interactions}
\label{subsubsec:qvcci}

The m331 model has eight vector bosons, the neutral ones: the photon, $A_\mu$, $Z_\mu$ and $Z^\prime_\mu$; and the charged ones: $W^\pm_\mu$, $V^\pm_\mu$ and $U^{\pm\pm}_\mu$.  
The Lagrangian terms for interactions among charged gauge bosons and quarks may be written as follows:
\begin{itemize}
\item $D$-type and $U$-type quarks interact as in the SM:
\begin{equation}
\mathcal{L}_{Wq}=\frac{g}{\sqrt{2}}\overline{D_{Li}}\,\gamma^\mu
\left(\mathbf{V_{CKM}}\right)_{ij}U_{Lj}W^-_\mu+H.c,\quad i,j=1,2,3.
\label{ccw}
\end{equation}

\item $J$ and $U$-type quarks:
\begin{equation}
\mathcal{L}_{VJ}=\frac{g}{\sqrt{2}}\overline{J_{L}}\,\gamma^\mu
(\mathbf{V_L^U})_{3j}U_{Lj}V^{+}_\mu + H.c.
\label{ccv}
\end{equation}

\item $J$ and $D$-type quarks:
\begin{equation}
	\mathcal{L}_{JU}=\frac{g}{\sqrt{2}}\overline{J_{L}}\,\gamma^\mu
	(\mathbf{V_L^D})_{3j}D_{Lj}\mathcal{U}^{++}_\mu + H.c., \quad j=1,2,3
	\label{ccu2}
\end{equation}

\item $j$-type and $D$-type quarks:
\begin{equation}
\mathcal{L}_{jV}=-\frac{g}{\sqrt{2}}\overline{j_{Lm}}\,\gamma^\mu
(\mathbf{O^\dagger V_L^D})_{mj}D_{Lj}V^{-}_\mu + H.c.,\quad m=1,2,\; j=1,2,3.
\label{ccv2}
\end{equation}

\item $j$-type and $U$-type quarks:
\begin{equation}
\mathcal{L}_{jU}=\frac{g}{\sqrt{2}}\overline{j_{Lm}}\,\gamma^\mu
(\mathbf{O^\dagger V_L^U})_{mj}U_{Lj}\mathcal{U}^{--}_\mu + H.c.,\quad m=1,2,\; j=1,2,3.
\label{ccu}
\end{equation}

\end{itemize}

The model has neutral currents coupled to $Z^\prime$ which are proportional to~\cite{Pisano:1993dg} 	
\begin{eqnarray}
\mathcal{L}_{Z^\prime}&\propto&[\overline{U_L}\gamma^\mu\mathbf{ V^{U\dagger}_L Y^U_LV^U_L}U_L+\overline{D_L}\gamma^\mu\mathbf{V^D_LY^D_LV^D_L}D_L\nonumber\\&+&\bar{U}_R\gamma^\mu_LY^U_RU_R
+ \bar{D}_R\gamma^\mu_LY^D_RU_R]Z^\prime_\mu
\label{fcnc}
\end{eqnarray}
where we have defined	
\begin{equation}
Y^U_L=Y^D_L
\propto\textrm{Diag}(
1-2s^2_W,\,1-2s^2_W,\,-1),
\label{yuleft}
\end{equation}
then, there are FCNC in the left-handed interactions but 
the right-handed neutral currents with $Z^\prime$ are diagonal in the flavour space since in this case we have $Y^U_R\propto \textbf{1}$ and $Y^D_R\propto\textbf{1}$. 
	
We see that all the matrices, $\mathbf{V_{CKM}}$, $\mathbf{V^U_L}$ and  $\mathbf{V^D_L}$, also survive se\-pa\-ra\-te\-ly in the left-handed interactions in Eq.~(\ref{ccw}) - (\ref{fcnc}). Moreover, as we show above, in the Yukawa interactions in Eq.~(\ref{cc1}) the matrizes $\mathbf{V^{U}_R}$ and $\mathbf{V^{D}_R}$, appear separately. We see that the origin of the parameters in the quark sector is Eq.~(\ref{yukawam331q}) in which two Higgs triplets contribute the quark masses. 

All this serves, at least, to make evident the beautiful simplicity of the standard electroweak model. However, it is possible that this simplicity could be its greatest weakness when dealing with processes that do not fit in its context. 

We can wonder ourselves, what are the values of each entry of the matrices $V^{U,D}_{L,R}$? and in the lepton sector, as we will se below, $V^\nu_L, V^l_L$ and $V^l_R$. How can we calculate something with so many parameters? A possibility is the following. Let's go back to the Eq.~(\ref{vdvup1})
and calculate $V^{U,D}_{L,R}$ solving simultaneously the following equations	
\begin{eqnarray}
&&V^{D\dagger}_LM^DM^{D\dagger}V^D_L=
V^{D\dagger}_RM^{D\dagger}M^DV^D_R=(\hat{M}^D)^2,\nonumber \\&&
V^{U\dagger}_LM^UM^{U\dagger}V^U_L=
V^{U\dagger}_RM^{U\dagger}M^UV^U_R=(\hat{M}^U)^2,
\label{vdvu2}
\end{eqnarray}
where $\hat{M}^U$ and $\hat{M}^D$ are defined in Eq.~(\ref{mumd}),
and at the same time using the numerical values of the quark masses, the CKM matrix, and the Yukawa couplings, $G,\tilde{G},F\tilde{F}$ in Eq.~(\ref{massud}), as input. Knowing the values of all the entries of the matrices $V^{U,D}_{L,R}$, only the masses of the extra particles remain  free parameters. The problem with this approach is that there are many (realistic) solutions for the entries of the $V^{U,D}_{L,R}$ matrices.
	
\begin{center}
\begin{figure}[!h]
\includegraphics[width=12cm]{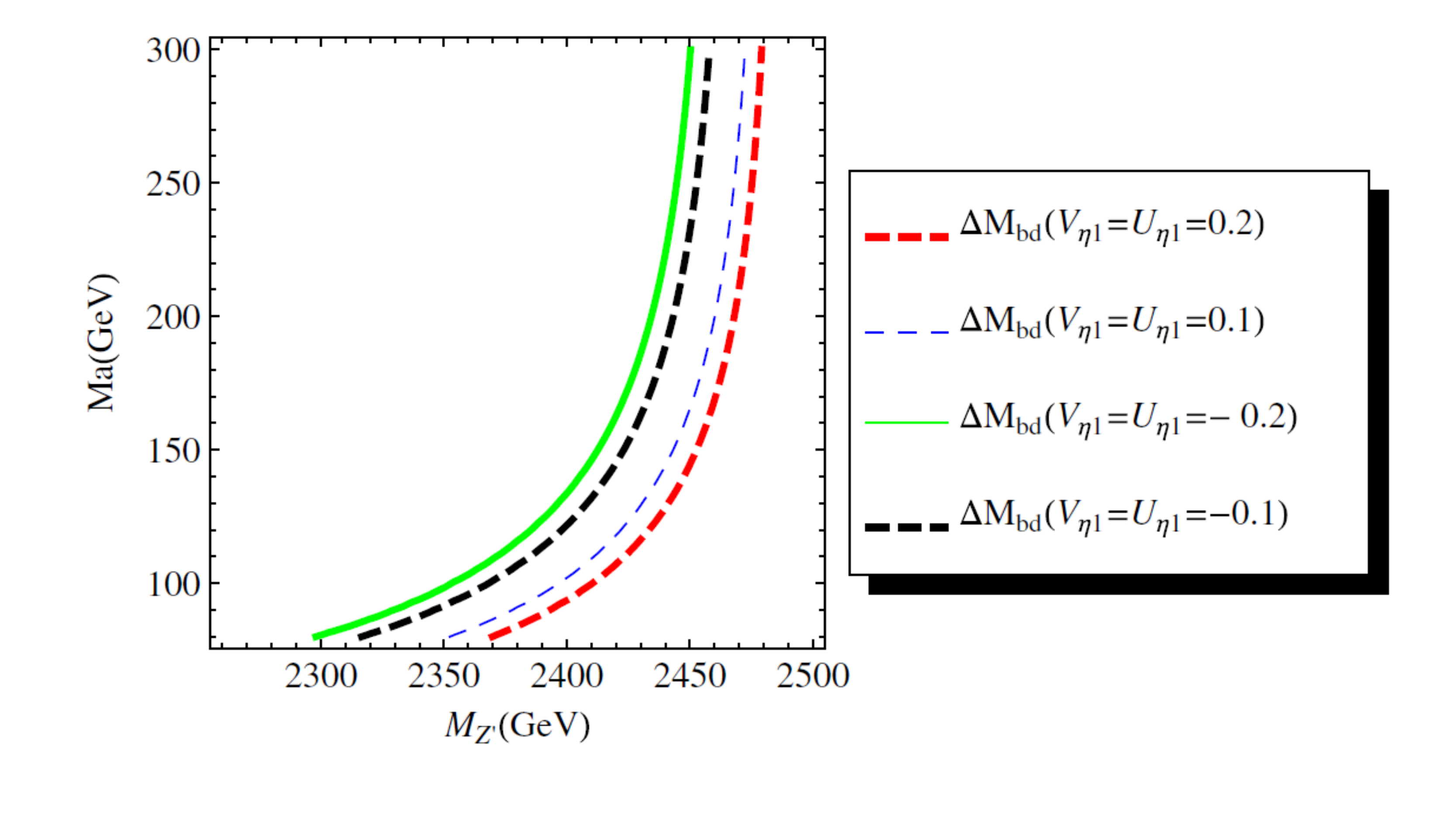}
\caption{Lower bounds on $\Delta M_M$ taken into account the contribution of $Z^{\prime}$ and a neutral scalar~~\cite{Machado:2013jca}. }
\label{fig:figura2}
\end{figure}
\end{center}

Using this method in the processes with $\Delta F=2$ and $\Delta F=1$, we note that, besides the FCNC via $Z^{\prime }$, the contributions of the neutral (pseudo) scalars may be important since, for some region of the parameter space, it may be negative interference. 
Taken into account only the contribution of $Z^{\prime}$ to the neutral mesons $\Delta M_M$ we have $M_{Z^{\prime}}>5$ TeV. However, by considering also the pseudoscalar contributions it is possible to obtain $M_{Z^{\prime}}\stackrel{>}{\sim}2.3$ TeV~\cite{Machado:2013jca}. See Fig.~\ref{fig:figura2}.

Another possibility, complementary to the previous one, is the following: we can consider the matrices $V^{U,D}_{L,R}$, and the masses of the extra particles (scalars,vectors and fermions), as free parameters. Then, as it was done in the SM, after a lot of phenomenology  using known and rare decays (possible anomalies), $\mathbf{(g-2)}_\mu$, ...  it will be possible to determine all the entries of the mixing matrices $V^{U,D}_{L,R}$, but without forgetting that, unlike the SM case, in the end, these matrices have to diagonalize the respective mass matrix. Of course, we can begin in a model- independent way by considering most known baryon and meson decays, and other processes, then continue with all the possible flavor anomalies.
In this case we can get rid of one the matrices, $V^U_L$ or $V^D_L$, by expressing one of them in terms of $V_{CKM}$ and the other one~\cite{Buras:2021rdg}. However, the matrices $V^U_R$ and $V^D_R$ still survive independently in the Lagrangian.

\subsection{Leptons}
\label{subsec:leptons}

Again, here we will consider only the m331 model~\cite{Pisano:1991ee,Foot:1992rh,Frampton:1992wt} in which the minimal version only includes the already known leptons.
As in the case of quarks, also in the leptonic sector the 3-3-1 models have many parameters. Although the interactions with $Z$ and $Z^\prime$ are diagoanl in the flavour space (the three generations of lepton transform in the same way under the electroweak symmetries $SU(3)_L\otimes U(1)_X$) there are FCNC in the scalar sector. The origin of the parameters is, as in the case of quarks, the different contributions to the mass matrices of charged leptons. If the masses are generated at the tree level, the Yukawa interactions include the triplet 
$\eta\sim(\textbf{3},0)$, in Sec.~\ref{subsec:quarks}, and the sextet $S\sim(\textbf{6},0)$ under the respective group factors. The components of the sextet are as follows
\begin{equation}
S=\left(
\begin{array}{ccc}
s^0_1 & \frac{s^-_1}{\sqrt{2}} & \frac{s^+_2}{\sqrt{2}}\\
\frac{s^-_1}{\sqrt{2}} & S^{--}_1 & \frac{s^0_2}{\sqrt{2}} \\
\frac{s^+_2}{\sqrt{2}} & \frac{s^0_2}{\sqrt{2}} & S^{++}_2
\end{array}
\right).
\label{sextet}
\end{equation} 

The Yukawa interactions in this model are give by
\begin{equation}
-2\mathcal{L}_{Yl}=\overline{(\psi_{aiL})^c}G^\eta_{ab}\epsilon_{ijk}\psi_{bjL}\eta_k+\overline{(\psi_{aiL})^c}G^S_{ab}\psi_{bjL}S_{ij}+H.c.
\label{lmasses}
\end{equation}
where $G^\eta(G^S)$ is a anti-symmetric (symmetric) matrix in the flavor space.

The scalar sextet in Eq.~(\ref{sextet}) has two neutral components, $s^0_1$ and $s^0_2$. If both gain a non zero VEV, $v_{s_1}$ and $v_{s_2}$ respectively, we obtain the lepton mass matrices
\begin{equation}
M^\nu=v_{s_1}G^S_{ab},\quad
M^l_{ab}=\frac{v_\eta}{\sqrt2}G^\eta_{ab}+\frac{v_{s_2}}{2}G^S_{ab},
\label{leptons2}
\end{equation}
it means that 
\begin{equation}
M^l=\frac{1}{\sqrt2}\left(v_\eta G^\eta+v_{s_2}\frac{M^\nu}{\sqrt{2}\,v_{s_1}}\right),
\label{lm1}
\end{equation}
and since $M^\nu/\sqrt{2}\,v_{s_1}\approx O(1)$, we see that $v_{s_2}> v_\eta |G^\eta|$
in order that the symmetric matrix could be the main contribution to the charged lepton masses. Notice that, since there are two source for the charged lepton masses, there are flavor changing neutral interactions in the scalar sector. We can not assume the charged lepton masses as being in some diagonal basis. This is a prediction of the model.

Next, let us consider the interactions in this sector sector.

\subsubsection{Lepton-neutral scalar interactions}

Omitting generation indices (scalar fields are still symmetry eigenstates)
\begin{eqnarray}
-\mathcal{L}^{nc}_{Yl}=\bar{l}_R(V^{l\dagger}_RG^\eta \eta^0+G^Ss^0_2)V^l_Ll^\prime_L+\overline{(\nu_L)^c}V^{\nu T}G^SV^\nu\nu_L s^0_1+H.c.
\label{intnl}
\end{eqnarray}
As in the quark sector we have to diagonalize the matrix $M^l$ in Eq.~(\ref{lm1}) by a bi-unitary transformation: 
$V^{l\dagger}_RM^lV^l_L=\hat{M}^l$ where $\hat{M}^l=\textrm{diag}(m_e,m_\mu,m_\tau)$; and
$V^{\nu T}M^\nu V^\nu=\hat{M}^\nu$~\cite{Bilenky:2020vjk}, with $\hat{M}^\nu=\textrm{diag}(m_1,m_2,m_3)$ in the neutrinos sector. The matrices $M^\nu$ and $M^l$ are given in Eq.~(\ref{leptons2}).
Hence, we have three irredutivel matrices: $V^l_L,V^l_R$ and $V^\nu$; or $V^l_R, V_L$ and introducing the PMNS matrix $V^\nu_L=V^l_LV_{PMNS}$. We stress that $V^l_L$ cannot be assumed as being the unity matrix. Below we use the first option i.e., the PMNS matrix appears in the lagrangian only when the right combination defined above appears in the interactions with the vector $W^\pm$.

\subsubsection{Lepton-charged scalar interactions}

These interactions are given by (charged scalar fields are still symmetry eigenstates)
\begin{eqnarray}
-\mathcal{L}^{cc}_{Yl}=\overline{l_{R}}
\left[(K^\eta_R)\eta^-_1+(K^S_R)\nu_{L}\frac{s^-_2}{\sqrt2}\right]\nu_{bL}+
\overline{(\nu_{L})^c}\left[(K^\eta_L)\eta^+_2
+
(K^S_L)\frac{s^+_1}{\sqrt2}\right]l_{L}
\label{intc1}
\end{eqnarray}
where we have omitted generation indices and defined $K^\eta_R=V^{l\dagger}_RG^\eta V^\nu$, $K^\eta_L=V^{\nu T}G^\eta V^l_L$; $K^S_R=V^{l\dagger}_RG^SV^\nu$ and $K^S_L=V^{\nu\,T}G^SV^l_L$.
As in the quark sector the matrices $V^l_{L,R}$ and $V^\nu$ survive separately in the Lagrangian. We are also assuming that there is no right-handed neutrinos.	

\subsubsection{Lepton-vector interactions}

In the lepton sector we have the interactions with singly charged vector bosons. The $\nu-l-W$ interaction 
\begin{equation}
\mathcal{L}_{\nu l W}= i\frac{g}{2\sqrt2}\left[\overline{\nu_L}\,\mathbf{e^{i\phi_l}\,
V^\dagger_{PMNS}}
\,\gamma^\mu l_L+
\overline{(l^c)_R}\,\mathbf{e^{^-i\phi_l}V^*_{PMNS}}\gamma^\mu (\nu^{c})_R\right] W^+_\mu+H.c.,
\label{nul}
\end{equation}
and the only difference with the SM is the phase $e^{i\phi_l}$, here as usual
\begin{equation}
V_{PMNS}=V^{l\dagger}_LV^\nu_L,
\label{pmns}
\end{equation}
and since neutrinos are Majorana particles, the $V_{PMNS}$ matrix have one Dirac and two Majorana phases.

There are also $\nu-l-V$ currents 
\begin{eqnarray}
\mathcal{L}_{\nu l V}= i\frac{g}{2\sqrt2}\left[\overline{(l^c)_L}\, \mathbf{V^*_{LR}}\gamma^\mu \nu_L 
-\overline{(\nu^c)_R}\,\mathbf{V_{LR}} \gamma^\mu l_R\right]V^+_\mu+H.c.
\label{barnul}
\end{eqnarray}
where we have defined
\begin{equation}
\mathbf{V_{LR}}=V^{\nu\dagger}_LV^{l*}_R.
\label{def}
\end{equation}
Hence, in the m331 model $V^\nu\not=V_{PMNS}$ always, and $V^l_L$ and $V^l_R$ appear in Eq.~(\ref{barnul}) in the combination showing in Eq.~(\ref{def}). We recall that $V^\nu$ appears also in the Yukawa interactions in Eq.~(\ref{intc1}). Notice that since the charged lepton mass matrix in Eq.~(\ref{lm1}) has a contribution of an anti-symmetric matrix, it is not possible to assume it diagonal from the very beginning, hence we cannot put $V^l_R=\mathbf{1}$ in the interactions in Eq.~(\ref{barnul}). This would be possible if the interactions leptons with the scalar triplet $\eta$  is forbidden, say by an appropriate discrete symmetry. However, in this case, the charged leptons mass matrix is diagonalized by the same unitary matriz that the neutrino mass matrix, and the PMNS is trivial. 

Then, as in the quark sector, it is necessary to known what are the values of the entries of the matrices $V^\nu$, $V^l_L$ and $V^l_R$ if we want to make a realistic phenomenology of this model.

It is possible to have mixing bewteem $W^\pm_\mu$ and $V^\pm_\mu$ (this occurs if neutral boson in the sextet $v_{s_1}\not=0$~\cite{Montero:1999mc}). In general, the global phase $e^{i\phi_l}$ in the charged current coupled with the $W^\pm_\mu$, see Eq.~(\ref{nul}), cannot be neglected anymore. The elimination of the global phase is possible only in the context of the SM where there is only one charged current coupled to $W^\pm_\mu$. In this case there is a contribution to the lepton electric dipole moment at the one-loop level. The same occurs in other models with several singly charged vector bosons as the left-right symmetric models~\cite{Chavez:2019yal}. 

In this model there are also interactions between charged leptons and doubly charged vector bosons given by the lagrangian
\begin{eqnarray} 
\mathcal{L}^U=
- \frac{g}{2\sqrt2}
\,\overline{l^c_a}\left[\gamma^\mu \mathbf{A_{ab}}-\gamma_5\gamma^\mu \mathbf{S_{ab}}\right]l_{b}\,U^{++}_\mu,
\label{u2+p}
\end{eqnarray}
$\mathbf{A}=\mathbf{\bar{V}}_{RL}-\mathbf{\bar{V}^T}_{RL}$ and $\mathbf{S}=\mathbf{\bar{V}}_{RL}+\mathbf{\bar{V}^T}_{RL}$ are antisymmetric and symmetric matrices, respectively; and
\begin{equation}
 \mathbf{\bar{V}}_{RL}=(V^l_R)^TV^l_L,
\label{def2}
\end{equation}

Finally, leptons couple universally with neutral vector bosons. For instance $Z^\prime$-lepton interactions
$Z^\prime$ is (univarsally) leptophobic the m331 model 
\begin{equation} 
f^l_V=-f^l_A=-\frac{\sqrt{3}}{6}\sqrt{1-4s^2_W}.
\label{fvfal}
\end{equation}
which are valid in some conditions discussed in Sec.~\ref{sec:solucao}.

\section{Challenges in the quark sector}
\label{sec:chaquarks}

Here, we will place some topics in the quark sector that in the future probably will deserve to be studied in more detail than what have been done so far. 

\subsection{In search of exotic quarks}
\label{subsec:exotics}

Flavor physics is capable of unveiling smaller distances than those that are possible in accelerators like the LHC. In fact, it may provide
an insight into scales as short as $10^{-21}$ m
(Zeptouniverse), corresponding to energy scale of
200 TeV or even shorter distance scales~\cite{Buras:2015nta}.

The search of vector-like singlet quark with exotic electric charge $-4/3,5/3$ has already been done in several colliders.  
When looking for this type of quarks, with electric charge $-4/3$, it is usually assumed $X_q\to b+W^+$ but data are consistent with the standard model (charge $q=+2/3$) and exclude the existence of an exotic quark with -$4/3$ electric charge and mass of the conventional top quark at the 99\% confidence level~\cite{cms43}. Quarks with charge $5/3$ has also been searched in processes as $X_{5/3}\to W^++ t$, being the $X_{5/3}$ mass, with right-handed (left-handed) couplings, below 1.32 (1.30) TeV excluded at 95\% confidence level~\cite{cms53}.
 
However, in the m331 model the transitions involving these exotic quarks are, see Eqs.~(\ref{ccv})-(\ref{ccu2}):
\begin{eqnarray} 
&&j_{-4/3}\to V^-+b,\quad j_{-4/3}\to U^{--}+t,\nonumber \\&&  
J_{5/3}\to V^++t,\quad J_{5/3}\to U^{++}+b. 
\label{ex1}
\end{eqnarray}
How important is having $V^-$ or $U^{--}$ instead of $W^-$? 

We would like to emphasize that, from D0~\cite{Abazov:2011xs} and CDF~ \cite{Aaltonen:2015xea}, to the LHC~\cite{CMS:2021mux,Sirunyan:2021sbg}, the search for $W^\prime$ has been done through the decay $W^\prime \to t+\bar{b}$. This is also the case in phenomenological analyses~\cite{Calabrese:2021lcz}. Of course, there are models in which this decay does exist.
However, in the m331 model the decays are $V^+\to j_{-4/3}+\bar{b}$ or  $V^+\to j_{5/3}+\bar{t}$. Moreover in Ref.~\cite{Calabrese:2021lcz} $W^\prime$ was considered that interacts with leptons with left-handed current, and the decay is $W^{\prime +}\to \tau^+_R+\nu_L$, but in the m331 the decay involving the extra charged vector boson is $V^+\to \tau^+_L+\nu^c_R$ i.e., through a right-handed current. 

\subsection{Do exotic bound states exist?}
\label{subsec:ebaryons}

The 3-3-1 models predict, under certain circumstances, the existence of new bound states that do not exist in the context of SM. For example, in the model with heavy leptons (331HL) the lightest of these leptons decay into the knwon leptons only throught a charged scalar  belonging to a triplet, $\eta^-_1$. 
Then, the lightest lepton may be stable enough to form atoms with one of the known charged leptons, the electron for example: $E^+e^-$. 

\begin{center}
\begin{figure}[!h]
\includegraphics[width=15cm]{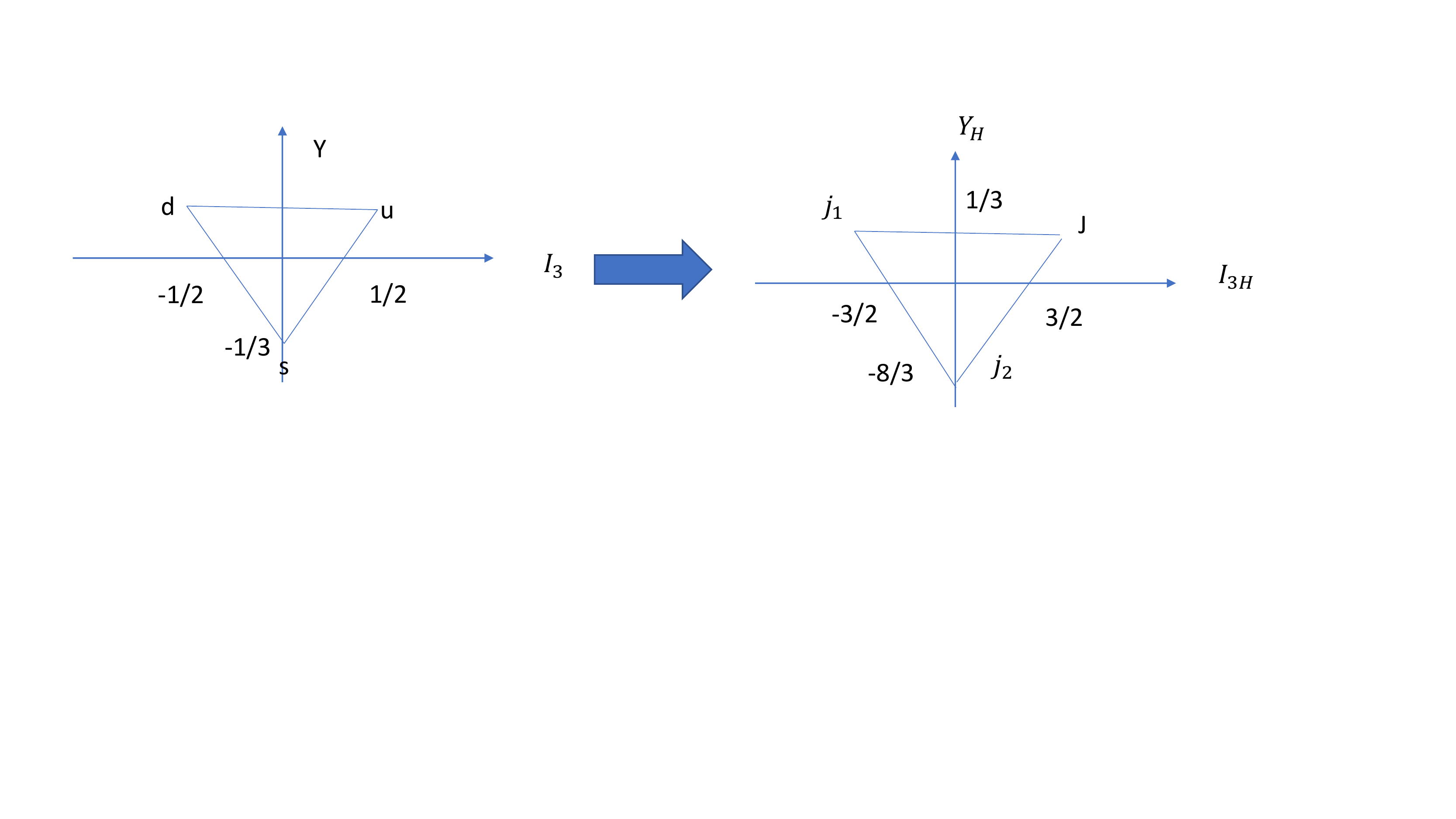}
\caption{Fundamental representation of the $SU(3)_H$. }
\label{fig:figura3}
\end{figure}
\end{center}

On the other hand, it has been observed that in the m331 and the 331HL models (both with $\beta=-\sqrt{3}$ and $X=0$)  there is one quark with electric charge $5/3$ and two with $-4/3$ (in units of $\vert e\vert$). Hence it is possible the existence of a replica of the  flavor $SU(3)_F$ symmetry  of the light quarks $u,d,s$ but now with the heavy quarks $J,j_1,j_2$ which 
would imply the existence of hadrons with exotic quarks~\cite{DeConto:2017aob}. It means that $SU(3)_F\to SU(3)_H$ making $u,d,s\rightarrow J,j_1,j_2$. See Fig.~\ref{fig:figura3}.

In this case, exotic hadrons are predicted with electric charge $\pm(3, 4, 5)$ with
\begin{equation}
Q=\kappa I_{3H}+\frac{Y_H}{2},\quad \kappa=3,\quad Y_H=B+S_H
\end{equation}	

Exotic baryons are for instance those in Tables~\ref{table1} and \ref{table2}.
Their decays are as follows:
$$
\Theta^{+5}_{3J}\to V^+ +\Theta^{+4}_{2Jt}\to 2l^++ 2\nu +\Theta^{+4}_{Jtt}\to ... 3l^++3\nu+3t\to...
$$	
	
Model with $\beta=-1/\sqrt{3}$ may have an $SU(3)_H$ with $u_4$ and $d_4,d_5$, then the electric charge are the same as the known mesons and baryons.

\begin{table}[ht]
\begin{tabular}{|l|l|l|} \hline  
Baryon              &            $Y_H$        & $Q/|e|$ \\ \hline
$\Theta_{3J}$       & $0$                     &  +5 \\ \hline
$\Theta_{2j_1j_2}$ &$-2$                     & $-4$  \\ \hline
\end{tabular}
\caption{Examples of exotic baryons in models with $\beta=-\sqrt{3}$. }
\label{table1}
\end{table}

\begin{table}[ht]
\begin{tabular}{|l|l|l|} \hline  
Meson &$Y_H$  & $Q/|e|$ \\ \hline
$\Pi^0_{\bar{J}J}$ & $1/3$  &  0 \\ \hline
$\Pi_{\bar{J}j_1}$ & $1/3$ & -3  \\ \hline
$\Pi^\prime_{\bar{J}j_2}$ & -3& -3\\ \hline
$\Pi^{\prime\prime}_{\bar{j}_1j_2}$ & -3&  0\\ \hline 
\end{tabular}
\caption{Examples of exotic mesons in models with $\beta=-\sqrt{3}$. }
\label{table2}
\end{table}

All these exotic hadrons may be long lived, with $\tau\sim 10^{-11} - 10^{-14}$ s (depending on the mass of the exotic quark and vector bosons).
Is it possible that these new degrees of freedom are waiting for us in a regime in which the theory is no more perturbative?

\section{Challenges in the lepton sector}
\label{sec:chaleptons}

Neutrino physics in 3-3-1 models has been widely studied in the literature. Here we only want to mention some aspects of this topic that we believe should be better studied in the future. 

\subsection{Neutrino masses}
\label{subsec:nusmasses}

The lepton mass matrices are given in Eq.~(\ref{leptons2}). We see that the charged lepton masses have contributions which are proportional to the neutrino masses. In fact, we can rewritte Eq.~(\ref{lm1}) in terms of the diagonal neutrino mass matrrix as follows:
\begin{equation}
	\hat{M}^l=V^{l\dagger}_L\left(\frac{v_\eta}{\sqrt{2}}G^\eta+\frac{v_{s_2}}{v_{s_1}}V^{l*}_LV^*_{PMNS}\hat{M}^\nu V^\dagger_{PMNS}V^{l\dagger}_L  \right)V^l_R,
	\label{lm2}
\end{equation}
where we have used $V^\nu= V^l_LV_{PMNS}$ and we see that it implies a fine-tuning. Moreover, if we use the Yukawa coupling in the matrices $G^\eta$ and $G^S$ and calculate the $V^l_{L,R}$ throuhg the charged lepton masses and $V^\nu$ with neutrino masses, it seems difficult to obtain at the same time the PMNS matrix~\cite{Machado:2018sfh}. 

If the scalar sextet is introduced, we have at least four possibilities: i) The fine tuning is possible and neutrinos are pure Majorana particles in this model, and no right-handed neutrino at all, at least for this issue. ii) $v_{s_1}=0$ and this condition is stable againts quantum fluctuations the Zee and/or Zee-Babu mechanims can be implemented and neutrinos gain masses at 1-loop or 2-loop order, respectivaly. Although the Zee mechanism was considered in Ref.~\cite{Okada:2015bxa} no conditions to maintain $v_{s_1}=0$ stable was given. Hence, in this case, neutrino masses are not calculables in the sense of Refs.~\cite{tHooft:1971qjg,Georgi:1972hy,Weinberg:1972ws,Georgi:1972mc}. In order to have calculable neutrino masses, we must protect $v_{s_1}=0$ against quantum effects while maintaining at the same time lepton number violation interactions in the scalar potential. iii) The other possibility is to add right-handed neutrinos implementing both type-I and type II seesaw mechanism. Finally, iv) to include more degrees of freedom as in Refs.~\cite{Kitabayashi:2001dg,Kitabayashi:2000nq,Kitabayashi:2001jp,Montero:2001ji}.

Which of these mechanisms are used in nature (if any)? 
In the first and second one neutrinos are pure Majorana particles and the PMNS matrix is exactly unitary. In the third case there must exist heavy right-handed neutrinos and the PMNS matrix is only approximately unitary, moreover the Keung-Senjanovic mechanism~\cite{Keung:1983uu} can occur. In the fourth one, the model is not the minimal version of the m331 since new degrees of freedom different from right-handed neutrinos have to be added.

Another possible way to generate lepton masses is the following: do not introduce the sextet or right-handed neutrinos. In this case the neutrinos are massles at tree level and charged leptons have the mass spectrum: $(0,m,m)$.  However, the Zee mechanism~\cite{Zee:1980ai} is still implemented by the $\eta^+_2$ scalar. It solves the problem with the neutrino masses but it is necessary still to obtain a way to give mass to electron and break out the mass degeneracy of the muon and tau including the PMNS matriz but using only the degrees of freedom of the m331. No such mechanism has been appear in the literature. But, we can wounder ourselves if is it possible a realistic mechanism for the charged lepton masses analogous to those proposed in Ref.~\cite{Georgi:1972hy}? 

Finally, we would like to comment an interesting possibility that was implemented in the context of the SM~~\cite{VanDong:2021xws} but which
may be implemented in 3-3-1 models with $\beta=-1/\sqrt{3}$ and other extensions of the SM. In this sort of models, the projection into the SM representation content introduce a vector doublet to the standard model, includes a vector doublet with $SU(3)_C\otimes SU(2)_L \otimes U(1)_Y$ quantum
numbers like the ordinary lepton doublets, $V_\mu=(V^0_\mu\,V^-_\mu)^T=(\textbf{1}; \textbf{2},-1/2)$. This vector doublet is suitable to address neutrino mass at the 1-loop level scotogenic mechanism because of the vertex $V^0_\mu V^{0\mu} H^0H^0$, where $H^0$ is the SM Higgs. The vector field is also dark matter, and contributes also to the muon anomalous magnetic moment. The dark matter candidate is either the lightest of the fermion singlets $N_{1;2;3}$, or one of the neutral vectors $V_1;V_2$ (mass eigenstates), that are stabilized by the accidental symmetry $\mathbb{Z}_2$. See also \cite{Saez:2018off}.

\subsection{Back to the old rare processes}
\label{subsec:rare}

It is interesting that most models with symmetry $SU(3)\otimes U(1)$ considered in Sec.~\ref{sec:interlude} were proposed by trying to explain the called ``three muon anomaly"~\cite{Benvenuti:1977fb,Holder:1977gp},
\begin{equation}
\nu_{\mu L}+N\to  \mu^+\mu^+\mu^- +X^-,
\label{3mu}
\end{equation}
and also the high-$y$ anomaly, which was an excess at high-$y$ in region, apparently observed in the process
\begin{equation}
\nu_\mu(\bar{\nu}_\mu)+N\to \mu^-(\mu^+)+X,
\label{highy}
\end{equation}
where $y=(E_\nu-E_\mu)/E_\nu\equiv E_h/E_\nu$ and $E_h$ is the hadron energy. Both processes were apparently
observed by the Harvard-Pennsylvania-Wisconsin-Fermilab (HPWF) collaboration~\cite{Aubert:1974en,Benvenuti:1977zp,Benvenuti:1976ad,Benvenuti:1976nq}.

Both processes cannot be explained in the context of the then called "Weinberg-Salam" electroweak model. 
New neutral leptons or doubly charged vector bosons? This was the context in which the old $SU(3)\otimes U(1)$ models considered in Sec.~\ref{sec:interlude}, were proposed.

\begin{center}
	\begin{figure}[!h]
		\includegraphics[width=9cm]{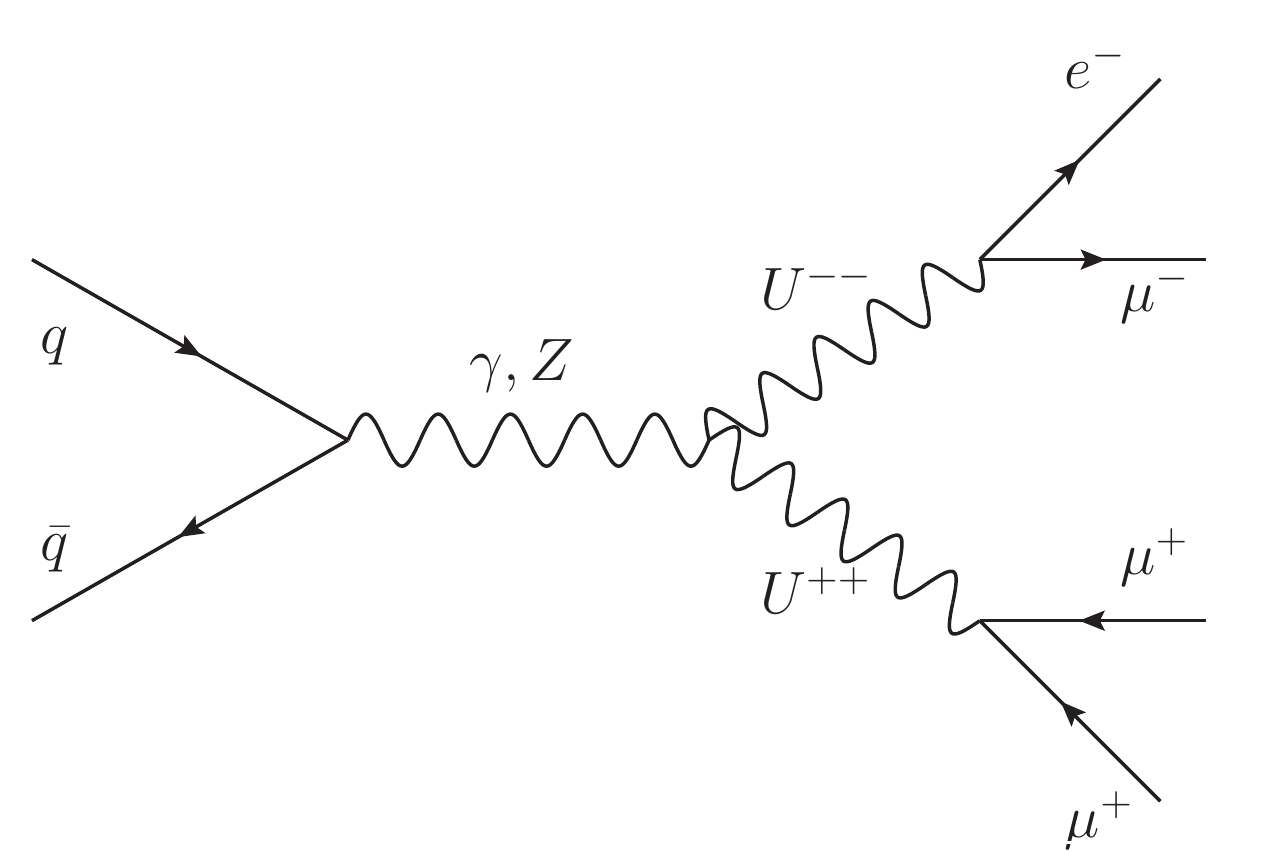}
		\includegraphics[width=10cm]{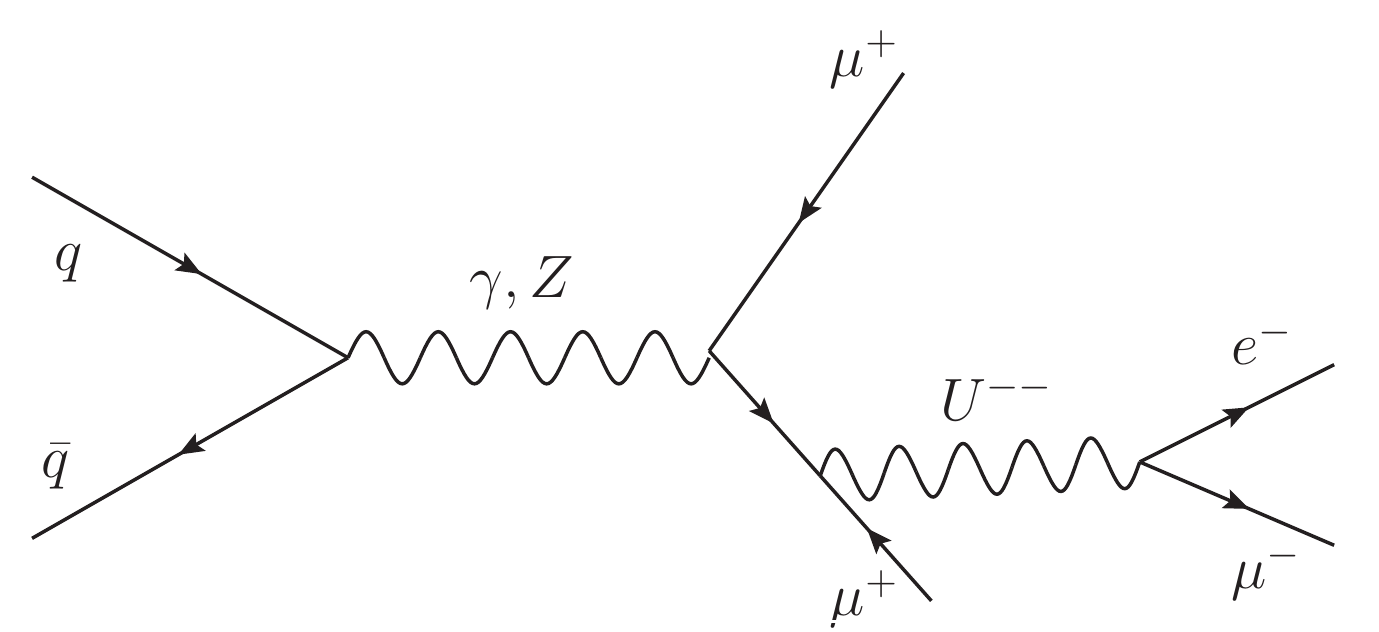}
		\caption{Three muons process at LHC. }
		\label{fig:figura4}
	\end{figure}
\end{center}

However, the Dortmund-Heidelberg-Saclay (CDHS) neutrino experiments at CERN 400 GeV SPS did not see neither the trimuons nor high-$y$ anomaly. Although no one ever knew what might be wrong with the HPWF experiments, since they were not confirmed by the CDHS collaboration, both processes were forgotten~\cite{Holder:1977en}. Should they?
	
\begin{center}
	\begin{figure}[!h]
		\includegraphics[width=14cm]{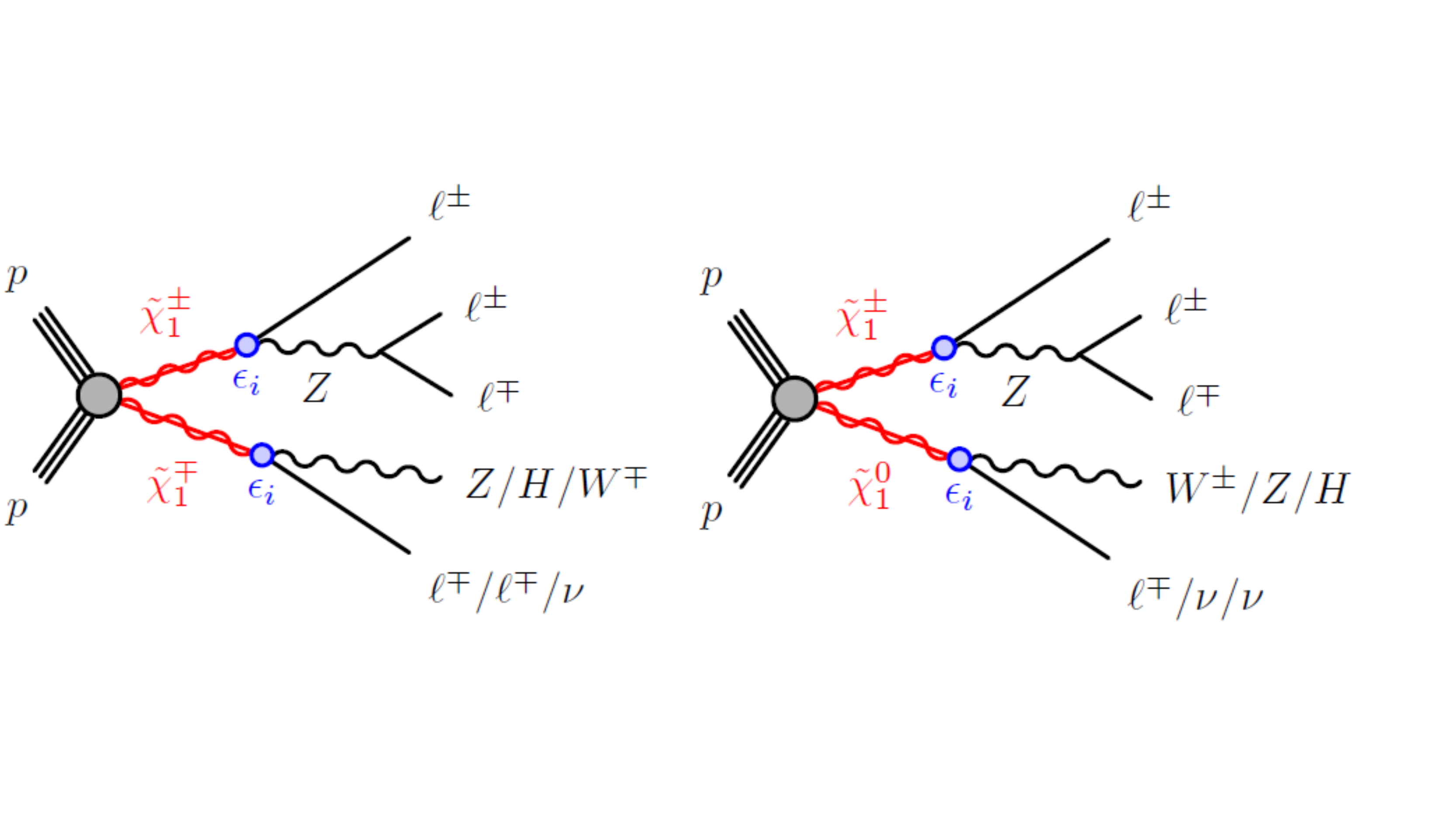}
		\caption{Three muons in susy models at LHC~\cite{Aad:2020cqu}. }
		\label{fig:figura5}
	\end{figure}
\end{center}

It happens that the three muon events occur in the m331 model, for instance at the LHC as can be seen in Fig.~\ref{fig:figura4} and also in supersymmetric models, through the production of a pair of chargino--neutralino also produce trimuon events in the final states~\cite{Aad:2020cqu,Aad:2021ajl}. 
See Fig.~\ref{fig:figura5}.

\begin{center}
\begin{figure}[!h]
\includegraphics[width=15cm]{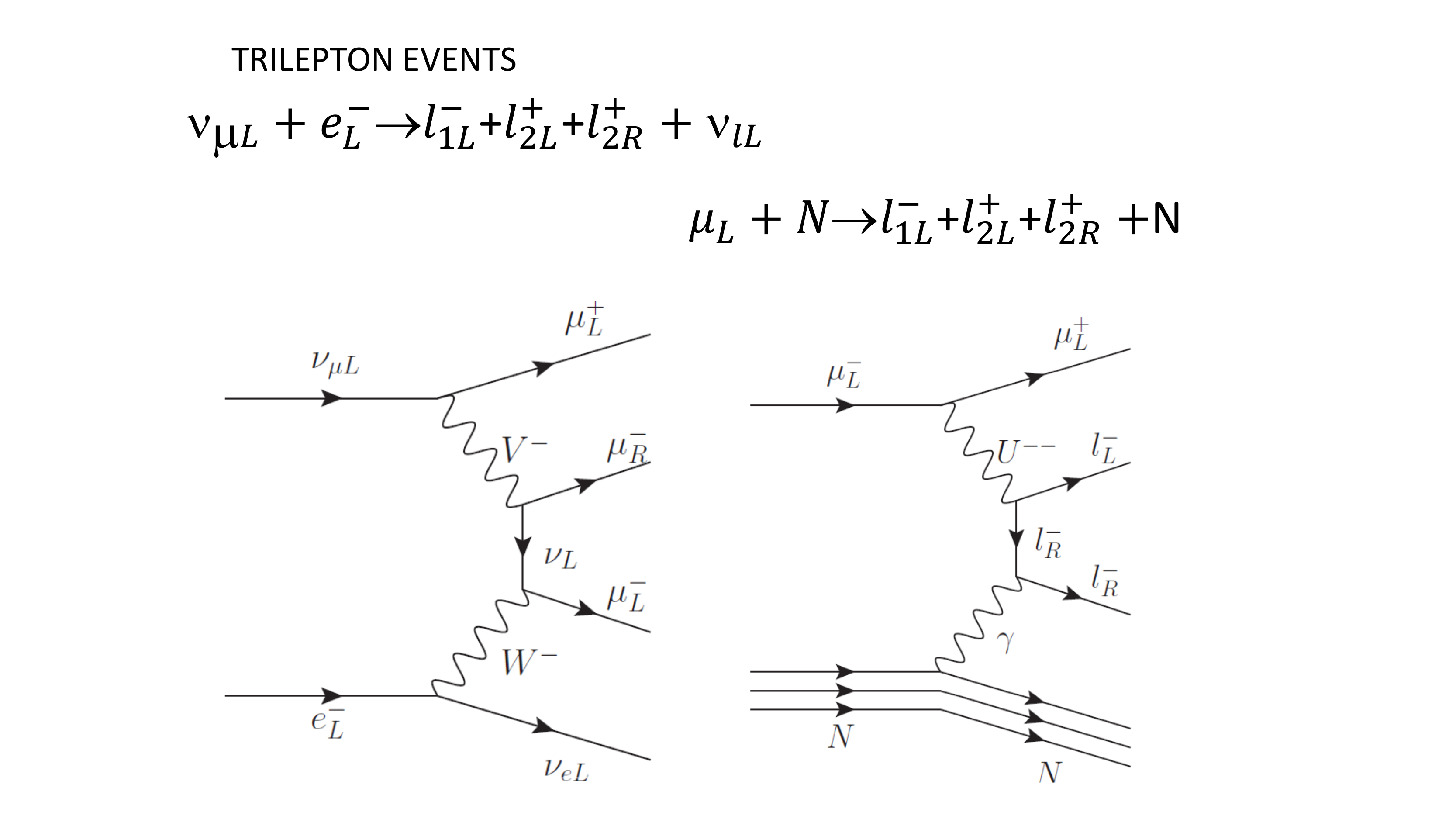}
\caption{Trileptons in $\nu_\mu-e$ and $\mu^--N$ scattering~\cite{Ge:2017poy}.}
\label{fig:figura6}
\end{figure}
\end{center}

Moreover, in the m331 model three leptons may also occur in $\nu_\mu-e$ and $\mu-N$ collisions as is shown in Fig.~\ref{fig:figura6}. 

\begin{center}
\begin{figure}[!h]
\includegraphics[width=14cm]{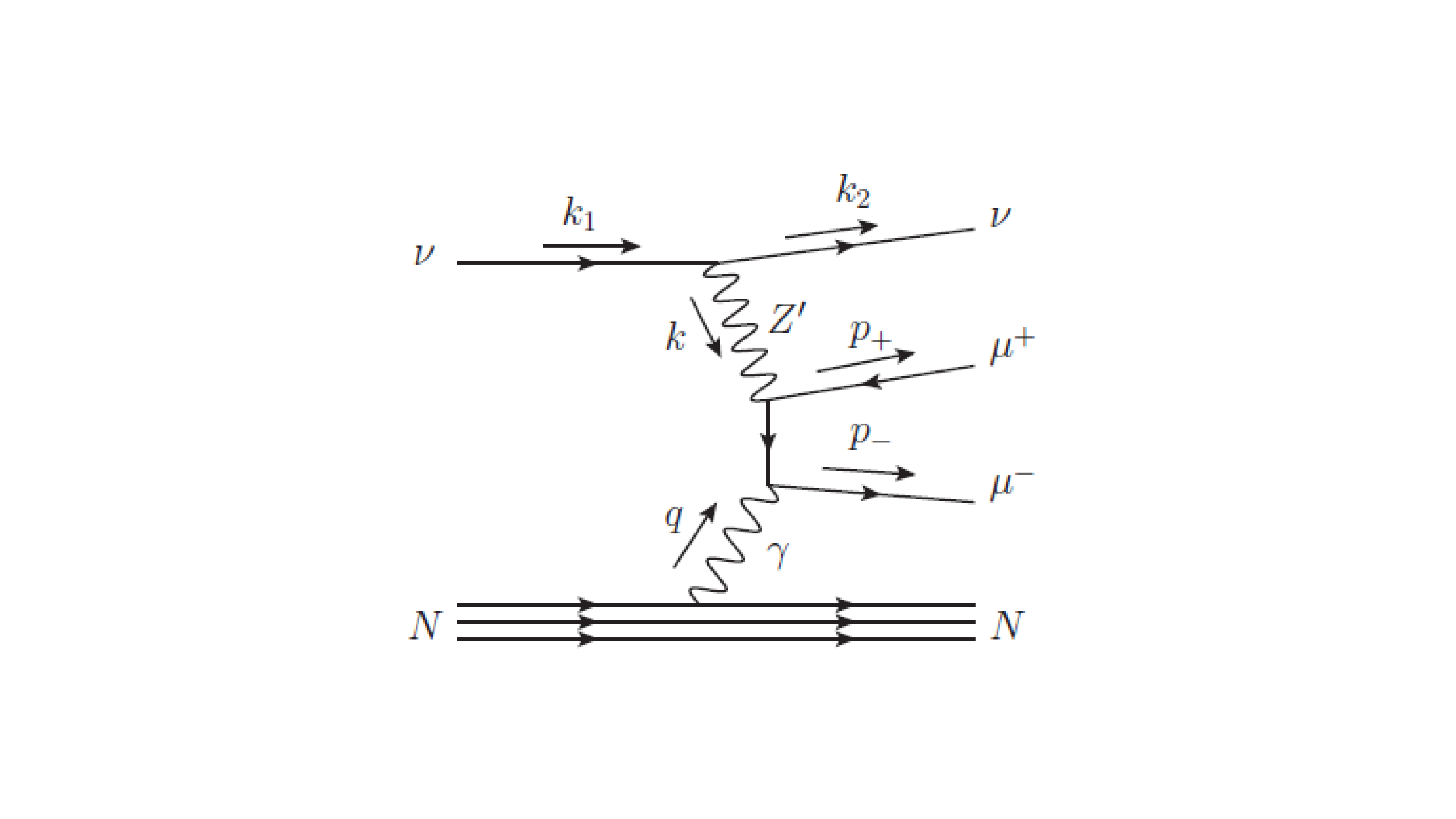}
\caption{Trident events in $\nu_\mu-N$ scattering~\cite{Altmannshofer:2014pba}.}
\label{fig:figura7}
\end{figure}
\end{center}

In 3-3-1 models (independently of the value of the parameter $\beta$) processes like the trident neutrino production~\cite{Altmannshofer:2014pba,Ge:2017poy}, shown in Fig.~\ref{fig:figura7}, may occur.

\begin{center}
	\begin{figure}[!h]
		\includegraphics[width=15cm]{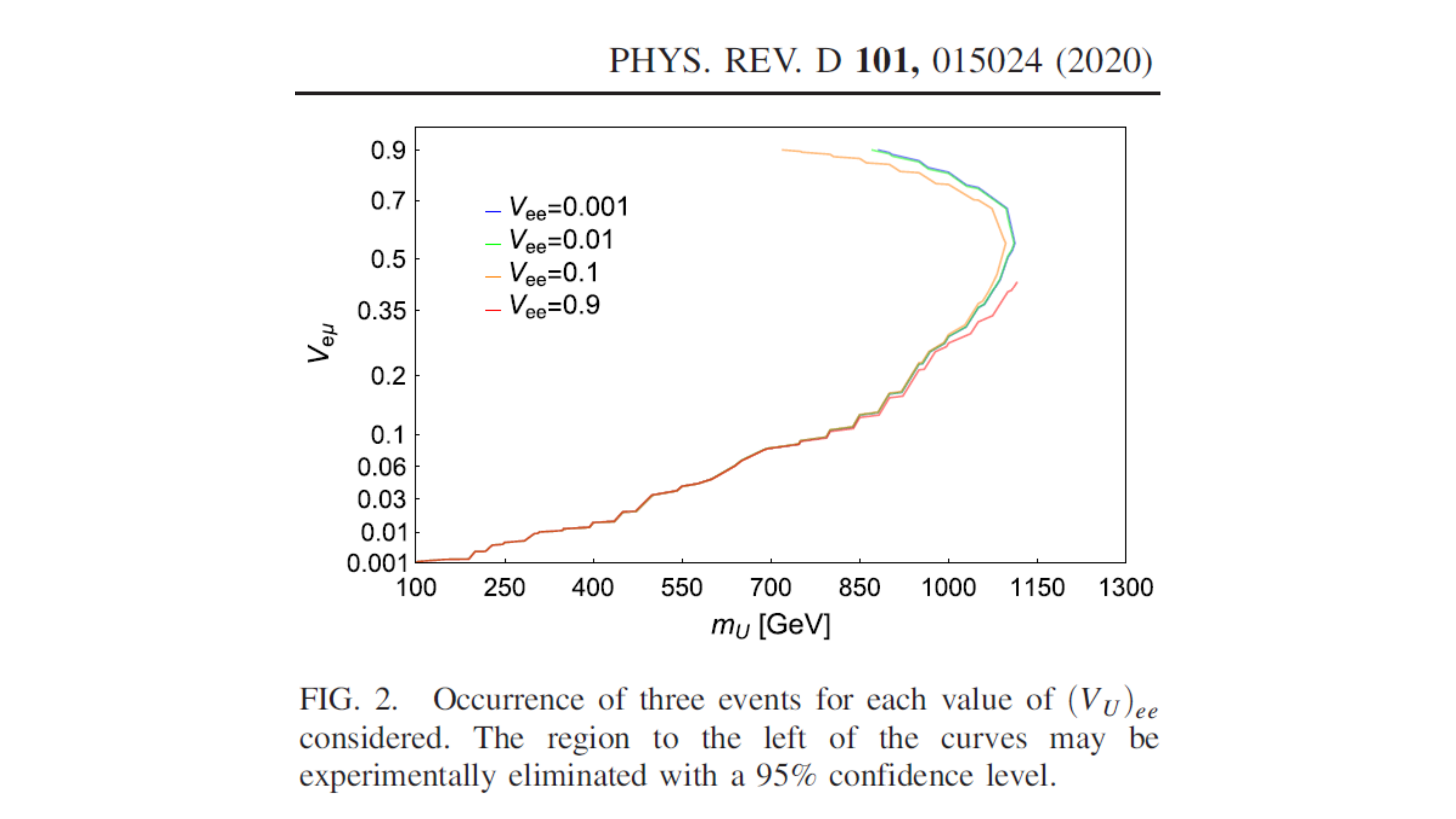}
		\caption{$pp\to 3\mu+X$ considering only SM+$U^{++}_\mu$. This process has no irreducible background since the SM vertices conserve flavor~\cite{Barela:2019pmo}.}
		\label{fig:figura8}
	\end{figure}
\end{center}

The process with three muons may occur at LHC also, and assuming the only source of this decay is the doubly charged vector
it could be discovery at the LHC if its mass is of the order of the 1 TeV~\cite{Barela:2019pmo}, see Fig.~\ref{fig:figura8}.

However the process $\mu\to 3e$ implies, under some reasonable assumptions, that~\cite{Machado:2016jzb}
$$M_{\mathcal{U}}>5,\;\;\textrm{ TeV}$$
In this case $U^{++}_\mu$ cannot be discovery at the present LHC.
However, this may not be the final word: both results can be made compatible if the contribution of the scalar fields with double charge were not negligible \textit{or/and} there are other realistic solutions for the leptonic mixing matrices. 
	
\begin{center}
	\begin{figure}[!h]
		\includegraphics[width=15cm]{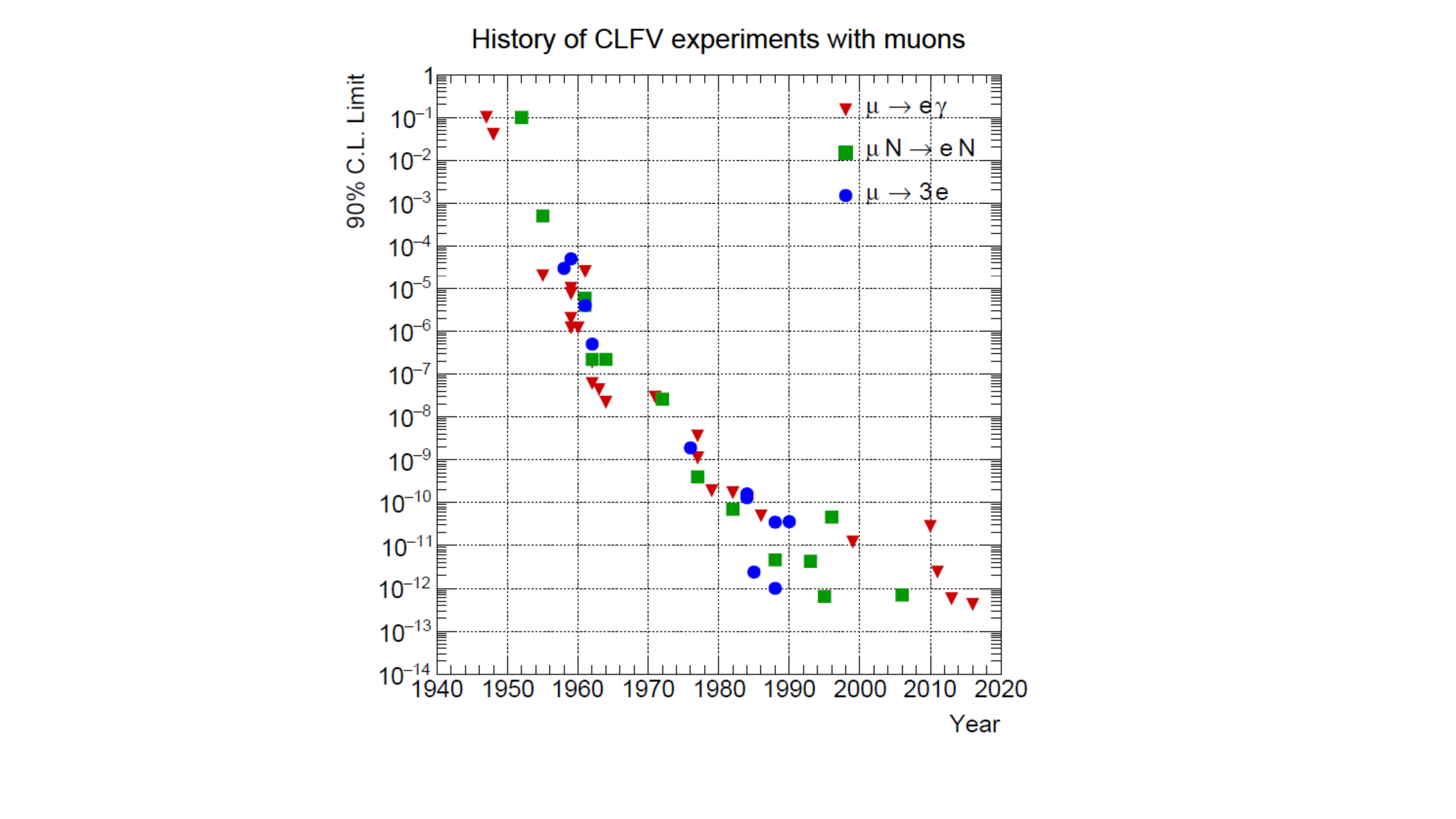}
		\caption{History of cLFV in the period 1940--2020~\cite{Galli:2019xop}. }
		\label{fig:figura9} 
	\end{figure}
\end{center}

In fact, charged lepton flavour violation (cLFV) processes are interesting for searching new physics due to the sensitivity of the next experiments, see  Table~\ref{table:tabela3} and Fig.~\ref{fig:figura9}. All models have to be confronted with the future data. This includes all 3-3-1 models. 

\begin{table}[!t]
\centering
\begin{tabular}{|c|c|c|}\hline
Decaimento &  Presente & Futuro\\
\hline
$\mu\to e\gamma   $ & $4.2\times10^{-13}$ & $6\times10^{-14} $ \\
$\mu N\to eN$  & $10^{-12}-10^{-13}$ & $10^{-17}$\\
$\mu\to ee\bar{e}$ & $10^{-12}$ & $10^{-16}$ \\ 
\hline
\end{tabular}
\caption{Lepton Flavour violation in muon decays, from~\cite{Galli:2019xop}.}
\label{table:tabela3}
\end{table}

It is possible to search for the high-$y$ and three muon anomalies, and similar processes, using the $\nu_\tau$ i.e., $\nu_\tau+N\to 3\tau+X$~\cite{SHIP:2021tpn}.

\subsection{Neutrino or antineutrino?}
\label{subsec:nusexp}

In all experiments with neutrinos it has been assumed, whether in the context of SM or beyond it, that neutrinos always accompany antileptons, while antineutrinos to leptons. 
In fact, this is true in the context of the SM where the only charged interactions are those coupled to the $W^\pm_\mu$~\cite{ParticleDataGroup:2020ssz}. 
Although, that in most extensions of the SM models having other charged interactions this continue to be true this is not the case in the m331 and probably in other models. The fact that the identification of the neutrino flavor cannot be done in a model-independent way, it was noted in Ref.~\cite{Grossman:1995wx}. Although, with the present precision of the experiments this does not seem necessary to be taken into account, it is conceptually correct. 

Let us remember that, in the SM, the leptonic transitions are the following:  
\begin{equation}
\quad W^+\to l^+_R+\nu_{lL},
\label{nusW}
\end{equation}
i.e., left-handed neutrinos of flavour $l$ are defined as the neutral particle emitted together with a right handed positron in the $ W ^ + $ decay,
while anti-neutrinos of flavour $l$ are defined as the neutral particle emitted together with a left-handed electron und a $W^-$:
\begin{equation}
 W^-\to l^-_L+ \nu^c_{lR}.
\label{antinusw}
\end{equation}
However, there may be situations where it appears some confusion in the identification of neutrinos and in some cases there is also a confusion in the definition of neutrinos and antineutrinos. Let us illustrate this situation considering the m331 in which there is a second singly charged vector boson.
In this model, besides the transitions in Eqs.~(\ref{nusW}) and (\ref{antinusw}), there are transitions as in Eq.~(\ref{barnul}),
\begin{equation} 
V^-\to  l^-_R+\nu_{lL},
\label{nusV1}
\end{equation}
which seems like an antineutrino not a neutrino, unless the chirality of the charged lepton is measured we cannot say that we are observing the process (\ref{nusW}) or (\ref{nusV1}). The transition 
\begin{equation} 
V^+\to (\nu^c_l)_R+l^+_L, 
\label{nusV2}
\end{equation} 
also occurs. Hence, a complete identification of neutrino or antineutrinos needs a measurement of the helicities of both leptons. 

Moreover, the vertices involving $V^\pm$ have a mixing matrix elements which are different from those in the usual PMNS matrix: $ \mathbf{V_{LR}\not= V_{PMNS}}$. See Eq.~(\ref{barnul}). There the mixing matrix is $\mathbf{V_{LR}=V^{\nu\dagger}_LV^{l*}_R}$ and, as we already dicussed, in the m331 it is not possible in any approximation $\mathbf{V^l_R=1}$. Hence any phenomenological analysis considering the interactions in Eq.~(\ref{barnul}) has to take into account the matrix $\mathbf{V_{LR}}$, otherwise, it cannot be interpret its result in terms of the m331. 

Although the confusion neutrino-anti-neutrino may not have a considerable effect in active neutrino oscillations, this is not necessarily the case with rare processes as neutrino$\leftrightarrow$anti-neutrino oscillations. This sort of oscillations may be viewed as the process 
$W^++l^-_L\to \nu_L\to (\nu^c)_R\to  l^+_R+W^-$
in which the intermediate-state neutrino propagates a macroscopic
distance~\cite{deGouvea:2002gf}. In this case the intermediate states requires a mass insertion $\nu\to \bar{\nu}$. In the m331 the respective process we have $l^-_L+W^+\to\nu_L\to l^+_L+ V^-$ (no helicity suppressed). Thus, in principle, both mechanism are distinguishable since the decays of $W^-$'s are different from those of $V^-$ and in the latter case there is no helicity suppression. For more details see Ref.~\cite{Pleitez:2022ido}.
		
\section{The SM non-trivial limit}
\label{sec:solucao}

We know that any model beyond the SM must, in some limit, coincide with the predictions of that model. This happens in particular with the 3-3-1 models with arbitrary $\beta$~\cite{Dias:2006ns}. Here we use as an example the m331 model. In this model the masses of the extra particles are proportional to $v_\chi$, the  VEV which breaks $SU(3)_L\otimes U(1)_X\to SU(2)_L\otimes U(1)_Y$, if they are fermions or, to $v^2_\chi$ if they are bosons. On the other hand, all particles that appear in the SM have masses which are proportional to a combination of the other VEVs, for instance $v^2_\eta+v_\rho^2+\cdots$ and so on. Hence it is obvious that for a large $v_\chi$, only the degrees of freedom whose masses are not proportional to $v_\chi$, or to $v^2_\chi$, will be excited at low energies.
For instance, the charged vector masses are given by
\begin{equation}
M^2_W=\frac{g^2}{4}v^2_W,\quad M^2_V=\frac{g^2}{4}(v^2_\rho+2v^2_{s_2}+v^2_\chi),\quad M^2_U=\frac{g^2}{4}(v^2_\eta+2v^2_{s_2}+v^2_\chi)
\label{mbvectors}
\end{equation}
where $v^2_W=v^2_\eta+v^2_\rho+2v^2_{s2}$ (neglecting the $v_{s_1}$ in the sextet).
All these models have two massive neutral vector bosons that we will denote
$Z_1$ and $Z_2$ and we will parameterize their neutral currents as follows~\cite{Zyla:2020zbs}:
\begin{equation}
\mathcal{L}^{NC}_{331}=-\frac{g_{3L}}{2}\sum_i\bar{\psi}_i\gamma^\mu[(\tilde{g}^i_V-\gamma_5\tilde{g}^i_A)Z_{1\mu}
+(f^i_V-\gamma_5f^i_A)Z_{2\mu}]\psi_i.
\label{nc1}
\end{equation}

Here, we consider only the couplings for leptons; for more details see Ref.~\cite{Dias:2006ns}. The couplings $g$'s  Eq.~(\ref{nc1}) are given for leptons by the exact expressions:
\begin{eqnarray}
&&\tilde{g}^\nu_V=\tilde{g}^\nu_A=-\frac{1}{3}N_1[1-6A(1+B)-3\bar{v}^2_\rho+4\bar{v}^2_W],\nonumber \\&&
\tilde{g}^l_V=-N_1(1-\bar{v}^ 2_\rho),\;\;\tilde{g}^l_A=\frac{1}{3}N_1[1-6A(1+B)-3\bar{v}^2_W],
\label{gvga}
\end{eqnarray}
where $\bar{v}^2_x=v^2_x/v^2_\chi$ with $v^2_x=v^2_\rho,v ^2_W$,  and $v^2_W=v^2_\eta+v^2_\rho+v^2_{s_2}$, being $v^2_W=(246\,\textrm{GeV})^2$, and $v_\eta,v_\rho,v_\chi$ are the vaccum expectation values for the triplets and $v_{s_2}$ for the sextet that are present in the model. Above we have defined
\begin{eqnarray}
&& N^{-2}_1=(1+4t^2_X)(\bar{v}^2_\rho-1)^2 +3\left[2(A(1+B)+\bar{v}^2_\rho-\frac{4}{3}\bar{v}^2_W-\frac{1}{3}\right]^2,\nonumber\\&&
N^{-2}_2=(1+4t^2_X)(\bar{v}^2_\rho-1)^2+ 3\left[2A(1-B)+\bar{v}^2_\rho-\frac{4}{3}\bar{v}^2_W-\frac{1}{3}\right]^2,
\label{def1}
\end{eqnarray}
and $A$ and $B$ are given by
\begin{eqnarray}
&& 3A=1+\bar{v}^2_W+3\left(1+\bar{v}^2_\rho\right)t^2_X,\nonumber \\&&
3A^2B^2=3A^2-(1+4t^2_X)\left[(1+\bar{v}^2_\rho)\bar{v}^2_W-\bar{v}^4_\rho\right]
\label{def3}
\end{eqnarray}
with $B>0$ and we have defined $t_X\equiv \tan\theta_X=g_X/g_{3L}$.

It was was pointed out in Ref.~\cite{Dias:2006ns} that, using all the exact expressions for the $M_{Z_1}$ mass in the ratio $M^2_{W}/M^2_{Z_1}c^2_W=1$ all neutral current couplings for quarks and leptons $g^i_{V,A}, i=q,l,\nu$ with $Z_1$ go to the tree level SM values, if we also use the matching condition $t^2_X=s^2_W/(1-4s^2_W)$. For instance, the complicate expressions in Eq.~(\ref{gvga}) for neutrinos become $g^\nu_V=g^\nu_A=1/2$. Similarly with the other fermions, the tree level expressions for $g^f_{V,A}$ are obtained and the $f$'s are only functions of $s^2_W$. For instance, $f^\nu_V=f^\nu_A=f^l_V=-f^l_A=-(\sqrt{3}/6)\sqrt{1-4s^2_W}$. The same happens with the model with $\beta=-1/\sqrt3$ but using $t^2_X=s^2_W/(1-\frac{4}{3}s^2_W)$.

Recently, it was shown that a lower bound on $v_\chi$ of 54 GeV can be obtained using only theoretical aspects of the model. This implies also a lower bound on the masses of the vector bileptons $M_U\geq 24$ GeV, $M_V\geq 78$ GeV and $M_{Z^\prime}\geq 78$ GeV~\cite{Barela:2021phu}. Of course, these are only theoretical values, phenomenology will say the last word.

A more realistic phenomenology could be done using the exact expressions in Eq.~(\ref{gvga}) and similar expressions for $f^i_{V,A}$. Then, we may see if more precise measurements shows how much the $Z_1$ could deviates itself from the predicted couplings of the $Z$ of the SM, including its mass.
After all, the only gauge couplings in the electroweak sector ar $g_X$ and $g_{3L}$. The standard model electroweak mixing angle $s_W$ appears only through out a matching (numerical) condition $t^2_X=s^2_W/(1-4s^2_W)$ at a given energy.
 
\section{Conclusions}
\label{sec:con}

Here we have shown that the parameter space of the 3-3-1 models is much wider than has usually been considered. In particular, we have stressed that the 3-3-1 models, since they have many parameters, need a more detailed study than  it has been done so far. 

We summarize the challenges of the 3-3-1 models as follows:
\begin{enumerate}
\item Search for exotic quarks $J,J_i$ (which are vector-like under the SM symmetries) at LHC and beyond~\cite{Cetinkaya:2020yjf}.
	
\item Direct, and indirect, search of the vector bosons $U^{\pm\pm}_\mu$ (models with $\beta=-\sqrt{3}$) or $\mathcal{Z}^0_\mu\not=\mathcal{Z}^{0*}_\mu$ (models with $\beta=-1/\sqrt{3}$). 
\item Are bounds on the mass of $U^{\pm\pm}_\mu$ using cLFV compatible with the discovery of this vectors at LHC? The answer seems to be yes~\cite{Barela:2022sbb}.
	
\item  Are right-handed neutrinos needed in the m331 model for generating neutrino masses? or, just as dark matter~\cite{Canetti:2012vf}? The model does not have a Majoron-like pseudoscalar, which that is part of the sextet, since $L$ is explicitly violated in some interactions in the scalar potential. Hence the mechanism of Ref.~\cite{Kelly:2021mcd} could be realized in the model.  
	
\item  In the model with $\beta=-\sqrt{3}$ and heavy charged leptons $E^\pm_l$ (which are vector-like under the SM symmetries) 
the production of four muons induced by the $Z^\prime$ in pp collisions are possible~\cite{Kawamura:2021ygg}. In the model with $\beta=-1/\sqrt{3}$ there are neutral vector-like leptons $N_l$, and this sort of process is also produced.
	
\item It is also important to continue the study of the effects of the charged scalars present in this models at the LHC~\cite{Coriano:2018coq,CMS:2021wlt} and other experiments. The m331 model has a doubly charged scalar field and, as all models with this sort of scalar, the process $S^{++}\to l^+l^+$ has signatures in $pp$ and lepton colliders, in particular in linear colliders~\cite{Bai:2021ony} and also effects in the $B$-meson decays~\cite{Cardozo:2020uol}. 
	
\item Study the possibility of these model under the constraints of long-standing discrepancy  of the $A^b_{FB}$ asymmetry with the standard model prediction~\cite{Yan:2021 veo}. For instance, the 3-3-1 models have many scalar multiplets and it is possible to accomodate the forward-backward anomaly in $Z\to \bar{b}b$ , as it has been done in the model with two and three Higgs doublets~\cite{Jurciukonis:2021wny}. 
	
\item Is it possible to have solution for the Cabibbo angle anomaly  \cite{Manzari:2021prf,Kirk:2021kcs} in these models? and also, is it possible to have combined solutions to the $A^b_{FB}$ asymmetry,
and some anomalous $B$ meson decays as in Ref.~\cite{Crivellin:2020oup} where vector-like quarks were introduced? In fact, models with $\beta=-1/\sqrt3$ allow the mixing $\overline{b_L(s)_L}\eta^{0*}D_R$ where $D_R$ is a vector-like quark with $Q=-1/3$ singlet under the SM group.   
	
\item The model with heavy leptons~\cite{Correia:2017vxa} deserves a more detailed study. As we said before, the lightest charged heavy leptons is almost stable since it decays into leptons only through charged scalars. In this model it is possible that the Higgs boson decay into long lived particles~\cite{ATLAS:2021jig,Alimena:2021mdu}. Is it possible the existence of heavy leptonic atoms, $E^+e^-$?
	
\item The phenomenology of the $Z^\prime$ at the energies of the FCC-ee~\cite{Oide:2021dye} should be considered, either in flavor physics ~\cite{Grossman:2021xfq}, or in forward-backward and left-right assymetries in lepton-lepton colliders~\cite{Montero:1998ve,Montero:2000ch,Montero:1998sv}. Would this $Z^\prime$ may be used to solve some $B$-mesons anomalies~\cite{Davighi:2021oel}. 
We must bear in mind that doing the phenomenology of the $Z^\prime$ independent of $\beta$~\cite{Binh:2021iww}, although it is an interesting step, does not consider, by principle, the contribution of the scalar fields whose representation content depends on the value of $\beta$.  
	
\item It is interesting to be able to distinguish physical fields in different representation, as in the scalar sector~\cite{Bandyopadhyay:2017klv}. 
	
\item Reliable values of quark and lepton masses and mixing matrix elements at a given energy scale are important for phenomenological studies at those energies. For instance, we can recognize the following energy scales (in GeV): $M_Z \sim 91.2$, the LHC $\sim 14\times 10^3$, the type I seesaw scale $\Lambda_N\sim10^9$, the Pecci-Quinn $\Lambda_{PQ}\sim 10^{12}$ and the would-be GUT scale $\Lambda_{GUT}\sim 10^{16}$~\cite{Xing:2007fb,Aparisi:2021tym}.
	
\item Are the new hadronic contributions to the $(g-2)_\mu$ anomaly important at least in some regions of the parameter space? The heavy quarks appearing in all 3-3-1 models could contribute through their vacuum polarisation to this parameter~\cite{Kennedy:2021ysp}. Also, non-perturbative effects could be important. Taking into account only electroweak contributions it may be not enough to explain the muon $g-2$ anomaly, even if these contributions give a positive result.

\item Revisited the decays $b\to sl^+l^-,\;l=e,\mu$ \cite{Buras:2013dea} and $c\to u\nu\bar{\nu}$~\cite{Colangelo:2021myn}.

\item Finally, we would like to call the attention to the recent CDF II Collaboration result: the mass of the $W^+$ does not agree with the SM calculation within almost 7$\sigma$~\cite{CDF:2022hxs}.
In the m331 model when all neutral components of the sextet gain a non zero VEV, the mass matrix of the singly charged vector bosons are given, in the $(W^+\,V^+)$ basis~\cite{Montero:1999mc}. Is this mixing enough to explain the $W$ mass? after all, the scalar triplet extension of the SM can solve the issue~\cite{Kanemura:2022ahw}. Or, the model with $\beta=-\sqrt3$ and heavy vector-like (under the SM symmetries) leptons may also solve th $W$ mass anomaly?~\cite{Nagao:2022oin}.

\item Production of leptons $l=e,\mu,\tau$ in proton-proton scattering:
\begin{eqnarray}
&& pp\to l^-_R l^+_L+(\nu_L)^c \nu_L (\slashed{E})+jets,\quad (a)\nonumber \\ &&
pp\to l^-_Rl^+_Ll^+_R+\nu_L(\slashed{E})+jets,\quad (b)\nonumber \\&&
pp\to l^+_L l^-_R+(\nu_L)^c\nu_L(\slashed{E})+jets,\quad (c)\nonumber \\ &&
pp\to l^-l^+l^-l^++jets,\quad (d)
\label{pp331}
\end{eqnarray}
may also occur because the resonances involving the exotic quarks are long-lived, since they are protected by a $\mathbb{Z}_2$ symmetry which is broken only by scalar interactions. For instance, the process (\ref{pp331})(d) occurs through the subprocess $u\to U^{--}j\to U^{--}U^{++}u\to l^-l^-l^+l^+u$, and so on. 

\end{enumerate}

All issues above must be clarify if the model is taken as a well motivated extension of the SM.

Finally some general remarks. There is one aspect that distinguishes the m331 from the other 3-3-1 models: there are \textit{necessarily} neutral interactions which violate flavor via neutral scalars. This is because, as we said before, it is not possible to assume the mass matrix of the quarks and charged leptons  in a diagonal form from the very start. For this reason, the PMNS matrix is not determined only by the matrix relating the neutrinos symmetry eigenstates with the mass eigenstates.
In other words, it is not possible to consider neither $V^l_L=\mathbf{1}$ in the PMNS matrix, nor $V^l_R=\mathbf{1}$ in the interaction with the vector $V^+$ in Eq.~(\ref{barnul}), hence all vertices necessarily involve the matrix $V_{LR}$ defined in Eq.~(\ref{def}) nor $\mathbf{\bar{V}_{LR}}$ in Eq.~(\ref{def2}). The same happens with the CKM matrix. This certainly has consequences for other processes. For instance, an analysis of the possibility of this model to explain, say the muon $g-2$ factor, has to use the interactions with $V^+$ given in Eq.~(\ref{barnul}). Similarly, with the interactions with the doubly charged vector, $U^{++}$ in Eq.~(\ref{u2+p}), in which the vertices include the matrices $\mathbf{A}=\mathbf{\bar{V}}_{RL}-\mathbf{\bar{V}^T}_{RL}$ and $\mathbf{S}=\mathbf{\bar{V}}_{RL}+\mathbf{\bar{V}^T}_{RL}$ with $\mathbf{\bar{V}}_{RL}$ defined in Eq.~(\ref{def2}).

Another point that we must emphasize is the following.
The scalar contributions may be important and cannot be negelected \textit{a priori}~\cite{Machado:2013jca,Barela:2021pzv}. Last but not least, if a model has new hadronic states, their contributions to the $(g-2)_\mu$ may be important. Moreover, the scalar cotributions cannot be neglected in all the paraeter space. 
Even more important is to take into account the hadronic contributions to the muon factor due to the new quarks in the models. Hence, without taking into account all the possible contributions to the $g-2$ we cannot say if a model is dead or alive. From the experimental point of view, it is not enough to just look for the existence of a certain type of particle, if it is found, it must be distinguished to which model it belongs.  

We stress once more, a realistic study of 3-3-1 models must take into accountn the values of the entries of the unitary matrices and also on the mass of the  extra particles in 3-3-1 models. Otherwise, it is like thinking that the SM could have been tested and confirmed without taking into account the CKM matrix.

Similar features occurs in model with the gauge symmetries $SU(3)_C\otimes SU(4)_L\otimes U(1)_N$~\cite{Pisano:1994tf,Jaramillo:2011qu,Nisperuza:2009xm,Palcu:2009wc}, or in models with $SU(3)_C\otimes SU(5)_L\otimes U(1)_N$~\cite{Pleitez:1993vf,Pleitez:1994pu,Palcu:2021nzl}. In fact, if right-handed neutrino do really exist in nature the more interesting model will be those with 3-4-1 symmetry~\cite{Pisano:1994tf} in which the known leptons and the right-handed neutrinos are in the fundamental representation of $SU(4)$. 

After more than a decade of finding no evidence for new physics at the LHC, we may wonder if we are studying the various theoretical proposals in the right way. Both experimental and phenomenological analysis usually are performed in a model-independent way or with a simplified version of a given model. 

From the experimental point of view, the importance of the reinterpretation of the experimental searches and measurements at the LHC in terms of models for new physics has been considered recently. In fact, there is no doubt that only a small subset of the possible theories and parameter combinations have been tested both, theoretically and experimentally~\cite{LHCReinterpretationForum:2020xtr}.

For instance, the importance of reinterpreting of experimental data can be appreciated in the case of neutral heavy leptons (HNL), $N$. ATLAS has considered the trilepton produced by a single HNL which production is possible via mixing with an electron \textit{or} muon neutrino produced by an on-shell $W$ boson decay. The search sets constraints on
the mass of $N$ in the range 4.5–50 GeV and a mixing $\vert U_{\mu}\vert^2$ of the order of $10^{-6}$~\cite{ATLAS:2019kpx}. However, reinterpreting the ATLAS data by considering two $N$'s mixing with the three neutrinos and with nearly degenerate masses leads to an exclusion limit on the total mixing angle which can be up to 3 orders of magnitude weaker than the limits reported by ATLAS~\cite{Tastet:2021vwp}. The model with two heavy neutral leptons is not just an unnecessary complication of the model with a single $N$: they are sufficient to explain neutrino masses and oscillations as well as
the origin of the matter-antimatter asymmetry, which are some of the motivation for that sort of leptons. Besides, a third HNL can play the role of dark matter. Hence, a model with only a single $N$ and which does not have these properties, cannot be considered a realistic model.

Another example that the reinterpretation of data seems to be necessary is the following. For years the MiniBooNE had found more electron neutrinos than has been expected at low energies. We recall that MinoBooNE, as all experiments detecting neutrinos, in fact, detect the corresponding charged lepton. Then, it is worth noting that this result is consistent in energy and magnitude with the excess of $\bar{\nu}_e$ reported by the 
LSND~\cite{LSND:1995lje,LSND:1996ubh,LSND:1997vqj}, and the significance of the combined LSND and MiniBooNE excesses is $6.0\sigma$~\cite{MiniBooNE:2018esg}. Moreover, these experiments used different detectors, Liquid Scintillator Neutrino Detector in the LSND, and liquid argon time projection chamber in the MiniBooNE. However, MiniBooNE does not distinghuishes electron from photons, and the latter ones could indicate another type of particle. Recently,  MicroBooNE~\cite{MicroBooNE:2021zai,MicroBooNE:2021nxr} had eliminated the possibility of extra events involving photons.
Then, we can speculate that the difference in the detectors used can not explain that difference. Would it be an effect of new physics the explanation of the excess observed in the two experiments? 
For instance, in the context of the m331 model, it is possible that MiniBooNE and LSND  confused neutrinos with antineutrinos since in this model the latter would also accompany electrons, or vice versa. See Eqs.~(\ref{nusV1}) and (\ref{nusV2}).

We see that is not an exaggeration to say that ``the field of particle physics is at the crossroads"~\cite{Fischer:2021sqw}. In fact, it is increasingly evident that we do not know how to get to observe physics beyond the SM, although we know that it must exist because of the issues mentioned in the introduction. 
On one hand, while we know there is physics beyond the SM, we have no idea where, and how, it will be manifest itself. On the other hand, we are certain that this model lost its predictive capacity. Thus, we may wonder ourselves, what are the hints of the range of couplings and masses to look for that have not been explored so far? We do not believe that the SM will remain valid up to the Planck scale. If this were the case, all the issues that the model does not explain~\cite{Beacham:2019nyx} would a priori remain unsolved. 

For all that has been discussed above, in the context of the m331 model (but which may be valid for any other sort of models) we can say with confidence that until now neither experiments nor phenomenological studies are in a position to confirm or refute any model. Of course, both the model-dependent and the model-independent approaches are necessary and must be continued. Direct search for new particles with as little model bias as possible~\cite{Komiske:2021vym} seems an illusory strategy. We recall the neutral current discovery. In this case the role of the theoreticians was crucial~\cite{Galison:1982bp}.

It is necessary that the \textit{ceteris paribus} condition must be left clear from the beginning of a work, be it theoretical or experimental.

\acknowledgments

The author would like to thanks  FAPESP under the funding Grant No. 2014/19164-6.

\end{document}